\documentclass[a4paper, twocolumn, 10pt, aps, amsmath, amssymb, accepted=2023-02-19]{quantumarticle}

\pdfoutput=1
\usepackage{soul}
\usepackage{graphicx} 
\usepackage{dcolumn} 
\usepackage{bm} 
\usepackage{dsfont} 
\usepackage{xcolor}
\usepackage{physics}
\usepackage{hyperref}
\usepackage{float}

\newcommand{\rwth}{Institute for Quantum Information, RWTH Aachen University, D-52056 Aachen, Germany}
\newcommand{\fzj}{Peter Grünberg Institute, Theoretical Nanoelectronics, Forschungszentrum Jülich, D-52425 Jülich, Germany}
\newcommand{\rhoin}{\rho_{\mathrm{in}}}
\newcommand{\rhoout}{\rho_{\mathrm{out}}}
\newcommand{\rhotarg}{\rho_{\mathrm{targ}}}
\newcommand{\psiL}{\ket{\psi_{\mathrm{L}}}}
\newcommand{\zeroL}{\ket{0_{\mathrm{L}}}}
\newcommand{\oneL}{\ket{1_{\mathrm{L}}}}
\newcommand{\HL}{\mathcal{H}_{\mathrm{L}}}
\newcommand{\XL}{X_{\mathrm{L}}}
\newcommand{\ZL}{Z_{\mathrm{L}}}
\newcommand{\pL}{p_{\mathrm{L}}}
\newcommand{\pidle}{p_{\mathrm{i}}}
\newcommand{\pnetwork}{p_{\mathrm{n}}}

\begin{document}

\title{Quantum Error Correction with Quantum Autoencoders}

\author{David F. Locher} \email{d.locher@fz-juelich.de} \affiliation{\rwth} \affiliation{\fzj}
\author{Lorenzo Cardarelli} \email{l.cardarelli@fz-juelich.de} \affiliation{\rwth} \affiliation{\fzj}
\author{Markus Müller} \email{markus.mueller@fz-juelich.de} \affiliation{\rwth} \affiliation{\fzj}

\date{March 6, 2023}

\begin{abstract}
Active quantum error correction is a central ingredient to achieve robust quantum processors.
In this paper we investigate the potential of quantum machine learning for quantum error correction in a quantum memory.
Specifically, we demonstrate how quantum neural networks, in the form of quantum autoencoders, can be trained to learn optimal strategies for active detection and correction of errors, including spatially correlated computational errors as well as qubit losses.
We highlight that the denoising capabilities of quantum autoencoders are not limited to the protection of specific states but extend to the entire logical codespace.
We also show that quantum neural networks can be used to discover new logical encodings that are optimally adapted to the underlying noise.
Moreover, we find that, even in the presence of moderate noise in the quantum autoencoders themselves, they may still be successfully used to
perform beneficial quantum error correction and thereby extend the lifetime of a logical qubit.
\end{abstract}

\maketitle


\section{Introduction}

Experimental platforms for quantum information processing are unavoidably subject to noise, which can cause failures during quantum computations.
The operation of reliable large-scale quantum computers will require active quantum error correction (QEC) procedures in order to cope with errors that dynamically occur during storage and processing of quantum information \cite{devitt2013,nielsenChuang2010,terhal2015}.
Quantum error correction relies on redundant encoding of logical quantum information, e.g.~into specific multi-qubit states or bosonic modes \cite{terhal2015,terhal2020}. 
Standard qubit-based QEC protocols require measurements of qubits that are coupled to the encoded data, followed by real-time feedback operations.
Experimental realizations of quantum error correction have seen great progress and range from repetition \cite{cory1998,chiaverini2004,schindler2011} and error detection codes \cite{linke2017,andersen2020} to recent fault-tolerant implementations \cite{hilder2022,egan2021,ryananderson2021,abobeih2022}.
While  performing in-sequence measurements and real-time feedback is experimentally challenging, it has  been achieved in various hardware platforms and application contexts \cite{riebe2004,barrett2004,sayrin2011,riste2013}.
Repeated cycles of quantum error detection and correction are currently studied extensively with superconducting qubits \cite{andersen2020,krinner2022,marques2022,chen2021,zhao2022}.
Experimental demonstrations of QEC that include in-sequence measurements and real-time feedback have been achieved in nitrogen-vacancy centers \cite{abobeih2022,cramer2016}, superconducting qubits \cite{andersen2019,riste2020}, trapped-ion platforms \cite{ryananderson2021,negnevitsky2018} and bosonic qubits \cite{ofek2016,hu2019}.
The correction of errors on the encoded data requires suitable feedback operations based on the obtained measurement results.
This task is known as decoding and is a subject of ongoing research \cite{terhal2015} that includes also the application of classical neural networks \cite{torlai2017,liu2019,maskara2019,sweke2021}. 
To avoid the challenges posed by in-sequence measurements and feedback, self-correcting quantum memories are investigated \cite{brown2016} and protocols for autonomous corrections have been considered.
Previous works in the latter direction include measurement-free QEC \cite{pazsilva2010,crow2016,premakumar2020} or engineered dissipation \cite{kerckhoff2010,pastawski2011,kapit2016,reiter2017}, e.g.~in bosonic codes \cite{leghtas2013,leghtas2015,lihm2018,gertler2021}.
Moreover, quantum machine learning represents a promising approach towards realizations of autonomous QEC that we want to follow in this work.
The field of quantum machine learning is rapidly developing in several directions \cite{schuld2014,biamonte2017,dunjko2018} ranging from variational quantum algorithms such as feedforward quantum neural networks (QNNs) \cite{cerezo2021,mangini2021,beer2020,kristensen2021,torrontegui2019} to quantum associative memories \cite{ventura2000,rebentrost2018,fiorelli2022}.
In this work, we focus on a type of multi-layered feedforward QNNs, called quantum autoencoders (QAEs), which have been investigated theoretically for the compression of quantum data \cite{romero2017,lamata2018,ma2023oncompression,bravo2021,cao2021}.
Furthermore, QAEs have been proposed to denoise specific quantum states such as GHZ- or W-states \cite{bondarenko2020,achache2021,zhang2021}.
Compression of quantum data using QAEs has already been achieved in experiments using single photons \cite{pepper2019,huang2020} or superconducting qubits \cite{ding2019}.
In other works, certain types of quantum neural networks were proposed to find suitable encodings of quantum information into logical states that allow for hardware-specific noise to be corrected \cite{johnson2017,cong2019, cao2022quantumvariational}.

In this paper we employ quantum autoencoders to perform quantum error correction and explore their utility with a focus on a quantum memory setting.
In contrast to most previous works, we envisage QAEs as a flexible and powerful tool to denoise generic states from a logical codespace instead of stabilizing specific quantum states, thereby extending the lifetime of encoded information.
Differently from conventional quantum error correction protocols, QAEs are intended to perform the error correction autonomously, requiring neither in-sequence measurements nor classical processing for decoding and feedback.
We show how QAEs can be used to correct computational errors on given logical states and also qubit erasures, which can be induced by the loss of qubits or leakage processes.
Additionally, we show that QAEs are able to adopt correction strategies that are suited optimally for the noise, which the QNNs are trained for.
We furthermore set up and analyze QNNs that can be used to unveil novel logical encodings in an unsupervised manner and without a-priori knowledge about the noise structure.
The discovered encodings are optimally suited to protect quantum information against that specific noise.
We propose and show that these QNNs can be directly transformed into QAEs ready to perform QEC on the newly discovered states without the need to conduct further training.
Lastly, we probe the robustness of the networks when these are constituted by noisy gates.
Our results show that even in the presence of moderate levels of intrinsic noise, QAEs can be used for beneficial quantum error correction, to extend the lifetime of a logical qubit.


\section{Background on QEC and QNNs}\label{section:model}

In this section we briefly summarize some basic quantum error correction concepts, which will be useful for the later benchmark of our QAEs against standard QEC codes.
Moreover, we review a model of multi-layered feedforward quantum neural networks, known in the literature under the name dissipative quantum neural networks \cite{beer2020,sharma2022,beer2021}.
We then set up quantum autoencoders using this model and discuss how they can be used for QEC purposes.

\subsection{Quantum Error Correction}

To protect quantum information from errors it is necessary to encode the information redundantly using for instance particular entangled multi-qubit states \cite{nielsenChuang2010,lidar2013}.
For a single encoded qubit, logical states $\psiL = \alpha \zeroL + \beta \oneL$ belong to the codespace $\HL$ spanned by two basis states $\zeroL$ and $\oneL$.
Many quantum error-correcting codes are conveniently described in the stabilizer formalism, which uses operators instead of state vector amplitudes to efficiently describe quantum states \cite{gottesman1997}.
An $n$-qubit stabilizer state is defined as the common $+1$-eigenstate of an Abelian group containing $2^n$ elements.
This stabilizer group is a subgroup of the Pauli group.
Without loss of generality, we focus on $n$ physical qubits encoding a single logical qubit.
The $2$-dimensional logical codespace is defined by a stabilizer group that can be generated from $n-1$ group elements.

\begin{figure}
    \centering
    \includegraphics[width=\linewidth]{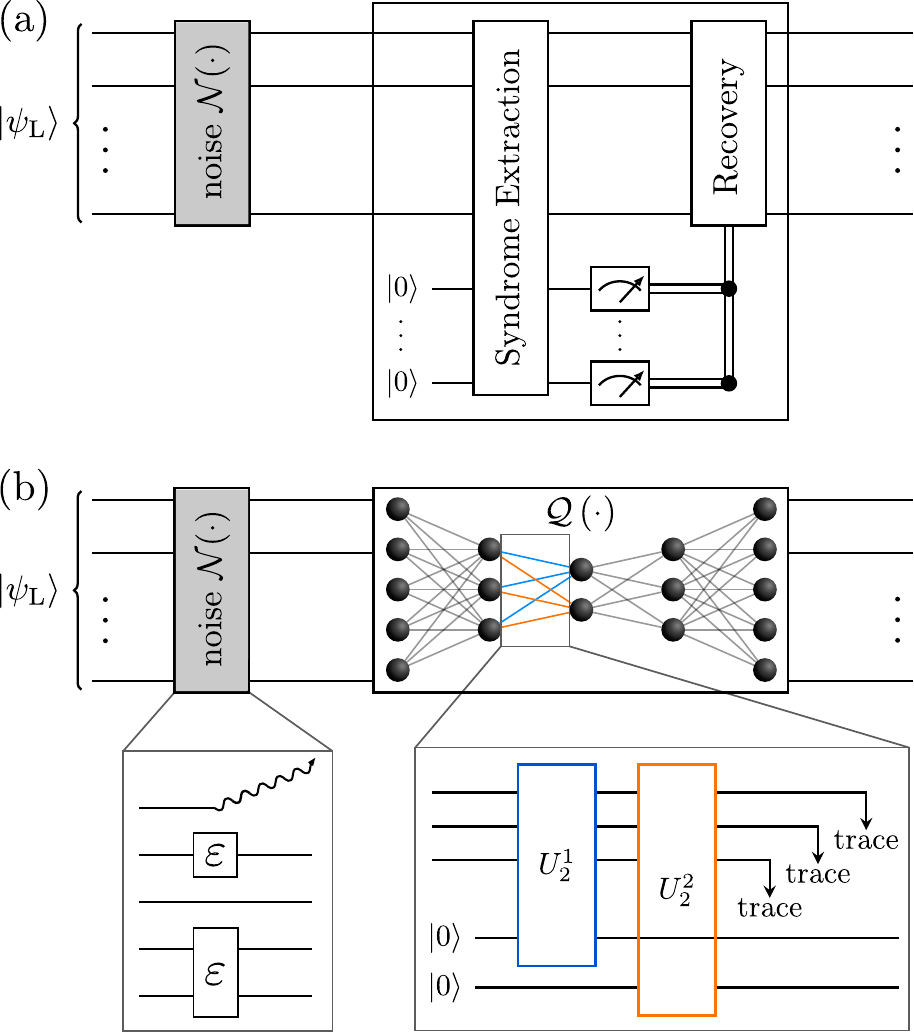}
    \caption{(a) Standard scheme of quantum error correction. Information is encoded in logical states $\psiL$ that undergo a noise process $\mathcal{N}$. Potential errors can possibly be detected by coupling ancilla qubits to the data and measuring the ancillas, yielding the error syndrome. Based on the syndrome an appropriate recovery operation is applied to the data qubits. (b) A quantum autoencoder being used for QEC. Instead of performing syndrome measurements and manually applying recovery operations we employ a QAE to perform the error correction autonomously. The network realizes a quantum channel $\mathcal{Q}$. We use the model of dissipative quantum neural networks to implement the QAE. Nodes in the graph represent individual qubits while edges represent unitary operations. The lower right box illustrates how layer-to-layer transitions are realized in a DQNN. Training the QAE amounts to learning the parameters of the unitary matrices. We find that QAEs can be successfully applied to correct computational errors as well as qubit erasures on logical states.}
    \label{fig:QAE_Intro}
\end{figure}

Errors occurring on the individual physical qubits can be detected if they map the state out of the logical codespace.
The correction of errors is conducted in two steps.
Measuring all stabilizer generators first determines for a possibly erroneous state the $2$-dimensional subspace, which is orthogonal to the original code space.
These measurements are performed by coupling ancilla qubits to the data qubits and measuring the ancillas, as depicted in Fig.~\ref{fig:QAE_Intro}(a) \cite{nielsenChuang2010}.
The measurements yield a set of $\pm 1$ outcomes that form the error syndrome.
A non-trivial syndrome indicates the occurrence of errors on the underlying logical state.
In a second step, potential errors must be removed, mapping the state back to the logical codespace $\HL$.
This is achieved by applying a suitable recovery operation to the data qubits being determined from the error syndrome in the process of decoding.
Different types of noise occurring on logical states may lead to different decoding strategies in order to achieve optimal error correction results \cite{terhal2015}.

\subsection{Dissipative Quantum Neural Networks}

A class of quantum feedforward neural networks having attracted attention in the last years are dissipative quantum neural networks (DQNNs) \cite{beer2020,bondarenko2020,beer2021,sharma2022}.
A DQNN can be represented as a graph consisting of neurons arranged in subsequent coupled layers, as depicted in Fig.~\ref{fig:QAE_Intro}(b).
A layer $k$ of the network consists of $n_k$ neurons that represent individual qubits.
The network as a whole realizes a quantum channel $\mathcal{Q}$ that maps an input state $\rhoin$ defined on the qubits of the first layer to an output state $\rhoout = \mathcal{Q}(\rhoin)$ on the last layer.
Each transition from a layer $k-1$ to a layer $k$ realizes an individual map $\mathcal{E}_k$.
The full network channel is the concatenation of all layer-to-layer maps:
\begin{equation}\label{eq:DQNNfeedforward}
	\mathcal{Q}(\rhoin) = \mathcal{E_\mathrm{out}} \left( \ldots \mathcal{E_\mathrm{3}} \left( \mathcal{E_\mathrm{2}} \left( \rhoin \right) \right) \ldots \right) .
\end{equation}
We regard the input layer as the first network layer, thus the layer-to-layer maps start at $\mathcal{E}_2$.
All constituent maps are adjustable via a finite set of parameters which can be chosen such that the network implements a map $\mathcal{Q}^{*}$ which achieves a desired task.
Training a network refers to the process of gradually adjusting the network parameters to eventually attain the target map.
Thus, DQNNs are set up similarly to classical feedforward neural networks \cite{goodfellow2016}, however, they implement quantum channels instead of maps on classical data.
The training of a DQNN is realized in a quantum-classical hybrid procedure: the network is implemented on actual quantum hardware while the optimization of the network parameters is performed on classical hardware.
For supervised learning, the training of a feedforward neural network requires training pairs in the form $\{ (\rhoin^{i}, \rhotarg^{i}) \}$, where states $\rhoin^{i}$ serve as input states for the network that one wants to be mapped to corresponding target states $\rhotarg^{i}$.
To quantify the success of the neural network in achieving this task, a cost function is defined which assumes its minimal value if the output states $\rhoout^{i} = \mathcal{Q}(\rhoin^{i})$ equal the corresponding target states.
A natural choice for a cost function is the averaged infidelity between training input and target states,
\begin{equation}
    C = 1 - \frac{1}{N} \sum_{i=1}^{N} \mathcal{F} (\rhoout^i, \rhotarg^i) ,
\end{equation}
where the fidelity between two quantum states $\rho_1$ and $\rho_2$ is defined as
\begin{equation}
\mathcal{F}(\rho_1,\rho_2) =  \left( \Tr \sqrt{ \sqrt{\rho_2} \rho_1 \sqrt{\rho_2} }\right)^2 .
\end{equation}
The quantum hardware is thus used to map training input states to output states, which are then measured to evaluate the cost function.
Classical optimization routines, such as widely used gradient descent algorithms, can then be used to find an updated set of network parameters that reduces the cost.
Updating the network parameters and repeating this cycle eventually leads to a convergence of the cost.
In this work we simulate the DQNNs on a classical computer.
This allows us to apply an efficient training algorithm similar to a backpropagation algorithm known from classical machine learning.
We sketch it briefly in Appendix~\ref{appendix:training_backpropagation} and refer to Ref. \cite{beer2020} for a detailed description.

\begin{figure}[t]
    \centering
    \includegraphics[width=\linewidth]{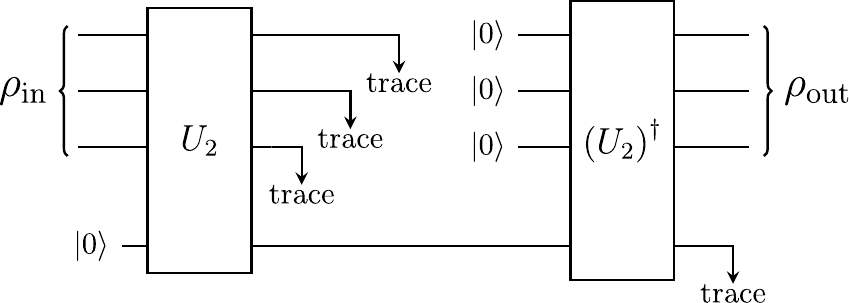}
    \caption{Quantum circuit that realizes a 3-1-3 QAE utilizing an architecture that we call \emph{self-inverse} architecture. Using this ansatz the decoding channel implementing the transition from the single-qubit hidden layer to the 3-qubit output layer is set up from the inverse of the unitary matrix implementing the encoding channel. Compared to independently trained channels this ansatz leads to a reduction of training parameters.}
    \label{fig:QAE_recurrent_architecture}
\end{figure}

We now describe how the layer-to-layer maps are realized in DQNNs.
The graph representation of a DQNN can straightforwardly be translated into a quantum circuit, as indicated in Fig.~\ref{fig:QAE_Intro}(b).
A layer-to-layer map $\mathcal{E}_k$ is implemented as follows.
Layer $k-1$ of the network, consisting of $n_{k-1}$ neurons, represents a quantum state $\rho_{k-1}$.
This state is supplemented with new qubits in the state $\ket{0}^{\otimes n_k}$.
A unitary matrix $U_k$ is then applied to the qubits of both adjacent layers.
Afterwards, qubits belonging to layer $k-1$ are discarded, resulting in a quantum state $\rho_k$ on the $k$-th layer of the network \cite{beer2020}:
\begin{equation}\label{eq:DQNN}
	\rho_{k} = \mathcal{E}_{k} \left( \rho_{k-1} \right) = \underset{k-1}{\Tr} \left[ U_{k} \left( \rho_{k-1} \otimes \ketbra{0}^{\otimes n_{k}} \right) U_{k}^{\dagger} \right] .
\end{equation}
The trace operation conducted in the maps gives rise to the term \emph{dissipative} QNNs.
The unitary operators $U_k$ mediating the layer-to-layer transitions are the trainable quantities in this model.
To reduce the number of training parameters or allow for an easy execution of the network map on actual hardware one may choose to set up the unitary matrices in various ways.
Beer \textit{et al.} \cite{beer2020} suggested to build an operator $U_k$ from $n_k$ individual unitary matrices $U_k^j$, each acting on all qubits in layer $k-1$ and a single qubit $j$ in layer $k$: $U_{k} = U_{k}^{n_k} \ldots U_{k}^{1}$.
This explicit realization is shown in Fig.~\ref{fig:QAE_Intro}(b) and we adopt this approach in our work.
In an experiment it is often more practical to  specify a parameterized ansatz for the unitaries, consisting of natively executable gates \cite{beer2021}, instead of training completely generic unitary matrices.

\subsection{Quantum Autoencoders}\label{subsection:QAEs}

Autoencoders (AEs) are technically defined as feedforward neural networks that are trained to reproduce their inputs at the output layer \cite{goodfellow2016}.
Typically, they comprise a hidden layer of width smaller than the input and output layers, meaning that some information must be discarded as the input states are processed.
Such networks are called undercomplete autoencoders.
Undercomplete AEs consist of two parts: a so-called encoder $\mathcal{E}$ maps the input to a latent state of smaller dimension.
A decoder\footnote{The term decoding in the context of autoencoders is not to be confused with the terminology of decoding as it is used in QEC.} $\mathcal{D}$ then tries to reconstruct the input from the latent state.
The full map $\mathcal{Q}$ describing the action of the autoencoder is thus a concatenation of the encoder and the decoder map: $\mathcal{Q}(\cdot) = \mathcal{D}(\mathcal{E}(\cdot))$.
Autoencoders can e.g.~be used for data compression \cite{tschannen2018}.
An undercomplete AE that succeeds to reproduce certain input data at the output layer is able to perform a lossless compression and reconstruction of the input.
The latent states can thus be considered as compressed data which are found in an unsupervised manner since no compressed reference states have to be provided for training.
Furthermore, AEs can be employed for denoising of data \cite{vincent2008}.
When an AE is trained to map noisy samples of the training data to noise-free instances, the network might learn to remove the noise and keep the relevant information while compressing the data.
Noise-free samples can then be reconstructed at the output layer.

Quantum autoencoders are defined equivalently to their classical counterparts: In the quantum case, the input and output states are quantum states and the network realizes a quantum map.
Just as classical AEs they can be split up into an encoding channel and a decoding channel, applied one after another.

In this work we employ denoising QAEs in the setting depicted in Fig.~\ref{fig:QAE_Intro}(b).
We consider arbitrary states $\psiL$ from a predefined codespace $\HL$ being affected by noise to serve as input states for a QAE.
We first assume that the network dynamics is noise-free, while in Sec.~\ref{section:internal_noise} we discuss the generalization to noise occurring during the application of the network.
The ultimate goal is that the QAE discards eventual errors while keeping the encoded quantum information as the noisy input states are processed.
To achieve this, we apply a supervised learning scheme using a small number of logical states $\ket{\psi^{i}_{\mathrm{L}}}$ from the code space.
We take noisy states $\mathcal{N}(\ketbra{\psi^{i}_{\mathrm{L}}})$ as training inputs and the corresponding noise-free logical states as target states for the training.
The cost function therefore reads
\begin{equation}\label{eq:costfunctionQAE}
	C = 1 - \frac{1}{N} \sum_{i = 1}^{N} \; \bra{\psi^{i}_{\mathrm{L}}} \rho_{\mathrm{out}}^{i} \ket{\psi^{i}_{\mathrm{L}}} ,
\end{equation}
with output states $\rhoout^i = \mathcal{Q}(\mathcal{N}(\ketbra{\psi^i_{\mathrm{L}}}))$.

In general, every layer-to-layer map of a DQNN is realized by independent unitary matrices that are adjusted during the training process.
However, the special form of a QAE consisting of an encoding and a decoding channel allows for a simpler ansatz.
Inspired by Ref. \cite{romero2017} we may choose to not train the encoding and decoding channels independently but set up the decoder using the inverse matrices from the encoder, as depicted in Fig.~\ref{fig:QAE_recurrent_architecture}.
In the remaining part of the paper we will refer to this ansatz as \emph{self-inverse architecture}.
The self-inverse architecture certainly leads to a reduction of training parameters and comes with additional advantages that will be highlighted in Sec.~\ref{section:QEC_results}.

\begin{figure}[t]
    \centering
    \includegraphics[width=0.99\linewidth]{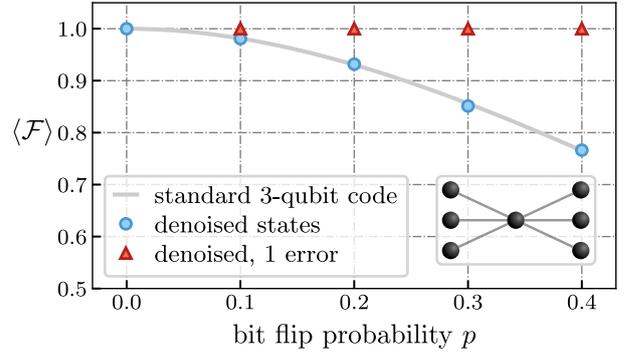}
    \caption{Validation of several 3-1-3 QAEs trained on logical states of the 3-qubit repetition code subjected to bit flip noise. To test the performance of a QAE, $10^4$ randomly drawn logical states are subjected to Pauli $X$-errors occurring independently on any qubit with probability $p$. The corresponding QAE having been trained on noise strength $p$ is then used to correct the errors on those states. The plot shows the averaged fidelity of denoised states w.r.t.~the corresponding noise-free logical states. We find that the QAEs (blue circles) perform exactly as well as the standard 3-qubit repetition code (grey line), $\langle \mathcal{F} \rangle = 1-\frac{2}{3}\pL$ (see Appendix~\ref{appendix:QPT}). A closer analysis reveals that the QAEs learn to perfectly correct single $X$-errors (red triangles). Here and in the remainder of the paper we omit error bars because the statistical errors are smaller than the symbol sizes. The inset shows a sketch of the 3-1-3 QAE geometry used in this example.}
    \label{fig:313_results}
\end{figure}


\section{Quantum Error Correction Results}\label{section:QEC_results}

In this section we present numerical results demonstrating that QAEs can be successfully used to perform quantum error correction.
In particular, we show that QAEs can correct both computational errors and qubit loss (quantum erasures) occurring on logical states of quantum error-correcting codes.

\begin{figure*}
    \centering
    \includegraphics[width=\textwidth]{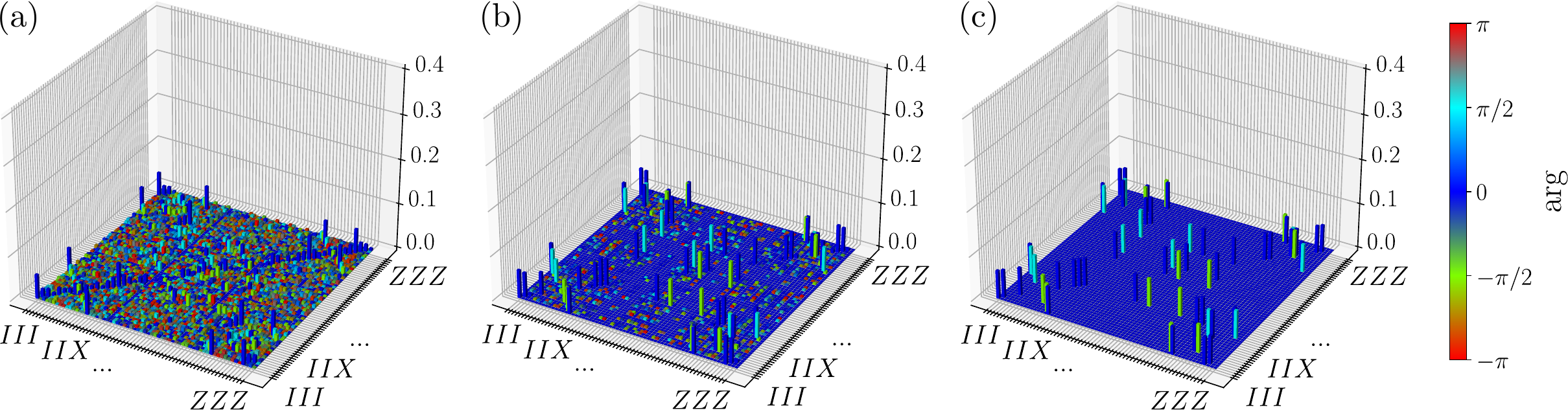}
    \caption{Quantum process tomographies of the maps which a 3-1-3 QAE, trained on bit flip noise of strength $p=0.1$, implements after 20 (a), 50 (b) and 100 (c) training epochs. As the training progresses, the quantum channel realized by the quantum neural network converges towards a final map which corresponds to the quantum process that realizes the correction map of the standard 3-qubit code. Details are given in Appendix~\ref{appendix:QPT}.}
    \label{fig:QPT_313}
\end{figure*}

\subsection{Correction of Computational Errors}

\subsubsection{Correction of bit flips on 3 qubits}

The 3-qubit repetition code (3QC) is a quantum error-correcting code able to correct single bit flip errors \cite{nielsenChuang2010}.
Here, we take the 3QC as an illustrative example to demonstrate that QAEs can be successfully trained and applied for QEC.
One may choose the logical codespace to be spanned by the states $\zeroL = \ket{000}$ and $\oneL = \ket{111}$.
This space is stabilized by a group generated e.g.~from the operators $Z_1 Z_2$ and $Z_2 Z_3$.
The assignment of logical basis states above fixes the logical generators of the code to be $\XL = X_1 X_2 X_3$ and $\ZL = Z_1 Z_2 Z_3$, up to multiplication by elements of the stabilizer group \cite{nielsenChuang2010,gottesman1997}.
A bit flip error happening to one of the physical qubits can be detected by measuring the two stabilizer generators.
Two bits of information allow for four different syndromes to be distinguished, corresponding to the noise-free case and the three different single-qubit $X$-errors.
Removing the respective bit flip corresponds to an appropriate recovery operation.
Bit flips occurring on two qubits simultaneously, e.g.~$X_1X_2$, are misinterpreted as single bit flip errors on the complementary qubit, $X_3$ for this example.
Therefore, a correction attempt causes a third bit flip, inducing a logical error $\XL = X_1 X_2 X_3$ on the state.
The presence of bit flip noise can be modelled via the bit flip channel which for a single qubit reads
\begin{equation}
    \mathcal{N}^{\mathrm{bit}}_{p}(\rho) = (1-p) \rho + p X \rho X .
\end{equation}
A 3QC state $\psiL$ that is subjected to independent bit flip noise suffers no error with probability $(1-p)^3$, a single bit flip with probability $3p(1-p)^2$ and two or three flips with probabilities $3p^2(1-p)$ and $p^3$, respectively \cite{nielsenChuang2010}.
Since single bit flip errors are correctable on logical states of the 3QC, active error 
correction on a noisy state induces a logical bit flip with probability $\pL = 3p^2(1-p) + p^3$ and recovers the noise-free state with prob. $1 - \pL$.

We start our study by training 3-1-3 QAEs, i.e.~quantum neural networks with a $3$-qubit input layer, a single-qubit hidden layer and a $3$-qubit output layer, on logical states of the 3QC.
A sketch of such a network is shown in Fig.~\ref{fig:313_results}.
For the training process we consider the three states $\zeroL$, $\oneL$ and $\ket{+_{\mathrm{L}}}$ which turn out to be enough training states for the network to learn to successfully generalize to arbitrary code states.
These states are subjected to bit flip noise occurring independently on the three physical qubits with probability $p$.
Concretely, we apply the corresponding noise channel to the states, meaning that mixed states $\mathcal{N}(\ketbra{\psi_{\mathrm{L}}})$ are taken as inputs for the QAEs.
As described in Sec.~\ref{subsection:QAEs}, the corresponding noise-free logical states are considered as target states.
As a benchmark, Fig.~\ref{fig:313_results} compares the trained QAEs and the standard 3QC in terms of the correction performance.
Every value of $p$ corresponds to a different QAE that has been trained on logical states subjected to noise of this strength.
A standard procedure in machine learning tasks is to test a trained neural network on data which has not been used for the training process, called validation.
Thus, for every trained QAE we randomly draw $10^4$ pure logical states that uniformly cover the logical Bloch sphere.
These states are subjected to Pauli $X$-errors occurring independently on every qubit with probability $p$ to then serve as validation input states for a QAE.
In an actual experiment, a network would be trained and run on the same platform, likely exposed to the same error conditions.
Hence, at both the training and testing stage we apply noise of equal type and strength.
At the validation stage we want to investigate how the QAE handles states that suffered no error, a single bit flip etc. Therefore, we use states $E_i \psiL$ as validation input states, where the set $\{E_i\}$ consists of all combinations of bit flip errors from which we draw errors with appropriate probabilities.
The averaged fidelity of denoised test states w.r.t.~the corresponding noise-free test states serves as a measure for the performance of the trained QAE.
For various noise strengths we find that the trained quantum networks perform just as well as the standard 3QC which corrects any single-qubit $X$-errors.
A closer analysis reveals that the networks for any $p < 0.5$ but $p \neq 0$ indeed learn to implement a channel that perfectly corrects any single bit flip error on arbitrary logical states, indicated by the red triangles in Fig.~\ref{fig:313_results}.
Hence, it is sufficient to train a QAE for one non-zero value of $p$ to learn a correction strategy that successfully generalizes to other noise strengths $p < 0.5$.
By performing quantum process tomographies of the maps $\mathcal{Q}$ realized by the fully trained QAEs, we are able to show that the learned channels are equivalent to the correction map of the standard 3-qubit code.
Figure~\ref{fig:QPT_313} shows quantum process tomographies of the maps that a 3-1-3 QAE implements at different stages of the training.
After $100$ training epochs there are no deviations of the quantum process from the map realizing the standard 3-qubit repetition code.
Details on the quantum process tomographies can be found in Appendix~\ref{appendix:QPT}.

\begin{figure}
    \centering
    \includegraphics[width=0.99\linewidth]{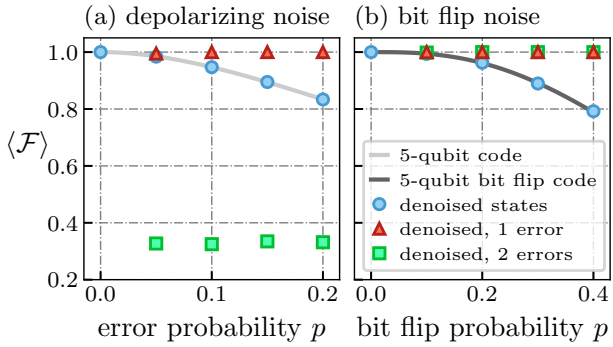}
    \caption{(a) Validation of several 5-1-5 QAEs trained on depolarizing noise. To test the performance of a network, $5 \times 10^4$ randomly drawn logical states of the 5-qubit code are subjected to Pauli errors occurring independently on every qubit with probability $p$. The QAE having been trained on the corresponding noise strength is then used to correct the errors on those states. The plot shows the averaged fidelity of denoised states w.r.t.~the noise-free validation states. We find that the QAEs, indicated by the blue circles, perform exactly as well as the standard 5-qubit error correction code represented by the grey line. A closer analysis reveals that the QAEs learn to perfectly correct single-qubit Pauli errors, as can be seen from the red triangles. (b) Validation of 5-1-5 QAEs trained on bit flip noise. Each network is tested on $10^4$ randomly drawn logical states subjected to independent $X$-errors. We find that the networks perform as well as a 5-qubit bit flip code, correcting all single- and two-qubit $X$-errors. Thus, we see that QAEs adopt different error correction strategies depending on the noise that is present during the training process.}
    \label{fig:515_results}
\end{figure}

We sometimes observe a failure of the training process, manifesting itself in a channel $\mathcal{Q}$ that maps arbitrary input states to a fixed state such as $\ket{000}$.
The training parameters realizing this map seem to correspond to a saddle point of the cost function, which is hard to escape using standard gradient descent methods.
However, employing the self-inverse architecture, shown in Fig.~\ref{fig:QAE_recurrent_architecture}, which reuses the unitary matrices from the encoding channel for the decoding channel
we do not observe these failures when training \mbox{3-1-3}~QAEs.

Considering QAEs whose latent space consists of a single qubit, it is clear that the first part of the network must conduct a combined correction and compression of erroneous logical states.
The network decoder then performs the trivial task of re-encoding a logical state.
Any errors left on the single-qubit intermediate state necessarily lead to logical errors on the final logical state.
However, there exists a gauge freedom in the sense that
the computational basis of the single-qubit state on the intermediate network layer can be arbitrarily rotated.
The network encoder can thus map an input state $\psiL$ to a state $R\ket{\psi}$ with $R$ being an arbitrary single-qubit rotation as long as the network decoder reconstructs the desired logical state $\psiL$ from $R\ket{\psi}$.
In Appendix~\ref{appendix:QPT} we visualize this feature by means of quantum process tomographies.

\subsubsection{Correction of arbitrary computational errors}

To be able to correct Pauli $X$-, $Y$- and $Z$-errors occurring on physical qubits in a quantum memory one has to employ an encoding that uses at least five qubits \cite{nielsenChuang2010}.
We consider logical states of the 5-qubit error-correcting code \cite{laflamme1996} being generated by the stabilizer elements
\begin{equation}
    \begin{aligned}
    g_1 = XZZXI \\
    g_2 = IXZZX \\
    g_3 = XIXZZ \\
    g_4 = ZXIXZ .
    \end{aligned}
\end{equation}
Logical states are $+1$-eigenstates of the operators $g_1$ to $g_4$ and the logical generators of the single encoded qubit may be chosen as follows:
\begin{equation}
    \XL = XXXXX , \quad \ZL = ZZZZZ .
\end{equation}
The 5-qubit code is the smallest distance-3 code which means that it can correct an arbitrary Pauli error happening to one of the physical qubits.
It satisfies the quantum Hamming bound: four stabilizer generators allow for $2^4=16$ different error syndromes to be distinguished, corresponding to the $15$ different single-qubit Pauli errors and the error-free case \cite{nielsenChuang2010}.

To further investigate the capabilities of quantum autoencoders to perform QEC we train QAEs with a 5-1-5 geometry and utilize them to correct errors on logical states of the 5-qubit code.
We set up the QAEs employing the self-inverse architecture introduced in Sec.~\ref{subsection:QAEs}.
Thus, the encoding channel is mediated via a single trainable 6-qubit unitary matrix while the decoding is realized using the inverse of that matrix.
As training input states we employ the six logical $X$-, $Y$- and $Z$-eigenstates of the 5-qubit code subjected to depolarizing noise.
The depolarizing channel for a single qubit reads
\begin{equation}
    \mathcal{N}_{p}^{\mathrm{depol}}(\rho) = (1-p) \rho + \frac{p}{3}(X \rho X +Y \rho Y + Z \rho Z)
\end{equation}
which we apply independently to the five physical qubits.
Fig.~\ref{fig:515_results}(a) shows a validation of the trained QAEs.
Every value of $p$ corresponds to a different QAE whose performance is tested by exposing it to randomly drawn logical states subjected to random Pauli errors according to independent depolarizing noise of strength $p$.
For various values of $p$ we find that the QAEs perform just as well as the standard 5-qubit error correction code that corrects arbitrary single-qubit Pauli errors.
Analyzing the action of those networks on different classes of errors exhibits that QAEs trained on $p \neq 0$ in fact implement a channel that perfectly corrects any single-qubit Pauli error.
Pauli errors of weight two are, however, not correctable.

\subsubsection{Adaptability to different types of noise}

To investigate whether QAEs can learn different error correction strategies in the presence of different types of noise, we consider again QAEs with a 5-1-5 geometry.
However, opposed to the previous example, we now train them on logical states that are subjected to solely bit flip noise.
As can be seen from a validation of these networks in Fig.~\ref{fig:515_results}(b), the QAEs perform just as well as a five-qubit bit flip code.
We find that the QAEs learn to perfectly correct up to two $X$-errors on arbitrary logical states.
This illustrates that QAEs can learn various error correction strategies depending on the noise suffered by logical states during the training process.

\begin{figure}
    \centering
    \includegraphics[width=0.97\linewidth]{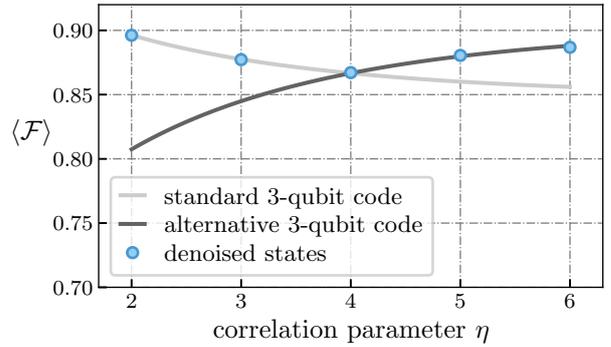}
    \caption{Validation of several 3-1-3 QAEs trained on logical states of the 3-qubit repetition code subjected to correlated bit flip noise with fixed overall bit flip probability $p=0.2$ and varying correlation strength $\eta$. For $\eta < \frac{1-p}{p}$ the probability for a single bit flip error to occur is larger than the probability of two bit flips happening on a state and vice versa for $\eta > \frac{1-p}{p}$. Thus, for the overall error rate $p=0.2$ the optimal error correction strategy is different for $\eta < 4$ and $\eta > 4$, denoted as standard / alternative 3-qubit code. QAEs automatically adopt the best possible denoising strategy during the training process. Each QAE is tested on $10^4$ randomly drawn validation states.}
    \label{fig:313_results_correlated}
\end{figure}

In experimental quantum information processing devices, noise can be correlated in space and time \cite{klesse2005,chubb2021}.
We want to study whether, in the presence of spatially correlated bit flip noise, QAEs can adopt correction strategies that perform better than the standard approach.
To do so, we go back to logical states of the 3-qubit repetition code, subjected to correlated bit flip noise.
We describe the noise by two parameters: an overall bit flip probability $p$ and a correlation parameter $\eta$ \cite{chubb2021}.
Choosing two qubits A and B, the correlation parameter is defined as
\begin{equation}
    \eta = \frac{\mathrm{Pr}(\mathrm{flip \ on \ A} | \mathrm{B \ flipped})}{\mathrm{Pr}(\mathrm{flip \ on \ A} | \mathrm{B \ not \ flipped})} .
\end{equation}
For simplicity, for any three qubits A, B and C we consider
\begin{equation}
    \begin{aligned}
        & \mathrm{Pr}(\mathrm{flip \ on \ A} | \mathrm{B \ flipped \ and \ C \ flipped}) \\
        = \; & \mathrm{Pr}(\mathrm{flip \ on \ A} | \mathrm{B \ flipped, \ C \ not \ flipped}) ,
    \end{aligned}
\end{equation}
i.e.~the probability for one of them to be flipped is the same, irrespective of whether one or both of the other qubits have suffered an error.
The case $\eta = 1$ corresponds to uncorrelated bit flip noise, whereas $\eta > 1$ describes bunching of errors, meaning that bit flips tend to occur in pairs.
Antibunching is characterized by $\eta < 1$.
We train 3-1-3 QAEs on logical states of the 3-qubit repetition code suffering bit flips with fixed probability $p=0.2$ but with different correlation strengths.
We find that networks trained in the presence of correlations $\eta > 4$ implement a different correction map than networks that were trained on small correlations.
This becomes clear from Fig.~\ref{fig:313_results_correlated}, where the performance of several trained QAEs is compared to the performance of the standard 3-qubit code and an alternative 3-qubit error correction strategy.
The alternative correction strategy considers flips of two qubits as most likely error events, therefore correcting those while inducing a logical error for single bit flips.
Indeed, if the correlation strength is increased beyond a certain threshold, it is advantageous to correct any two errors instead of single ones.
Abbreviating the conditional probability $\mathrm{Pr}(\mathrm{flip \ on \ A} | \mathrm{B \ flipped})$ as $p_{\mathrm{c}}$, the turnover point resides at $p_{\mathrm{c}} = 0.5$.
From Bayes' theorem follows that $\mathrm{Pr}(\mathrm{flip \ on \ A} | \mathrm{B \ not \ flipped}) = \frac{p(1-p_{\mathrm{c}})}{1-p}$, so we find the critical correlation strength to be $\eta_{\mathrm{c}} = \frac{1-p}{p}$.
During the training process, a QAE correctly determines and implements the correction strategy which yields the best error correction results.

These findings demonstrate that QAEs can adopt error correction strategies that are optimally suited for the type of noise present during training.
We saw this at the example of QAEs adapting to depolarizing and bit flip noise or QAEs adjusting their correction strategy in the presence of correlations.
In an experiment, a QAE would be trained on the device on which it is supposed to perform the error correction later on, thus implementing an optimal denoising strategy for the specific device.
In the canonical scheme of syndrome-based quantum error correction the different strategies which a QAE can embrace correspond to different decodings of syndromes into recovery operations.

\subsection{Correction of Erasures and Computational Errors}

Besides computational errors, losses and leakage errors pose a threat for successful quantum computations \cite{grassl1997}.
Ions or neutral atoms for example might escape the trapping potential in atomic systems \cite{stricker2020,baker2021} or photons can get lost from a photonic quantum processor \cite{lu2008}.
Moreover, leakage into states which are not part of the two-dimensional qubit subspace represents a risk, e.g.~in superconducting \cite{fowler2013,varbanov2020,battistel2021} or atomic \cite{brown2018} quantum processors.
The quantum erasure channel, which for a single qubit reads
\begin{equation}
    \mathcal{N}_{p}^{\mathrm{erasure}}(\rho) = (1-p) \rho + p \ketbra{2} ,
\end{equation}
is used frequently to model losses or incoherent leakage errors.
The positions of possible erasures can be detected in experiments by performing quantum non-demolition measurements.
These signal the occurrence of potential erasures while leaving the quantum state invariant if no erasures have happened.
Such detection protocols have been proposed or even implemented for various architectures \cite{stricker2020,varbanov2020,wu2022published}.
To be able to protect a logical qubit from single erasures, a code consisting of at least four physical qubits is necessary \cite{grassl1997}.
Moreover, any complete code of distance $d$ can be used to correct $d-1$ erasures or located errors \cite{grassl1997}.

\begin{figure}[b]
    \centering
    \includegraphics[width=\linewidth]{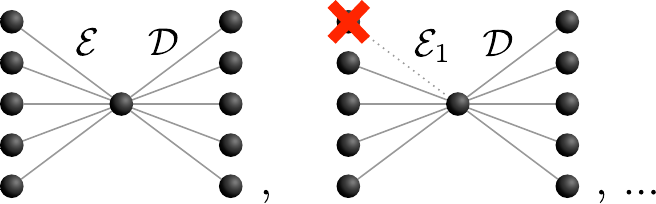}
    \caption{Collection of QAEs used to correct erasures of qubits. Since the positions of erasures are known, every possible combination of erasures is corrected by a separate QAE. States $\mathrm{Tr}_{1}\left(\ketbra{\psi_{\mathrm{L}}}\right)$ for example are processed by a QAE consisting of the channels $\mathcal{E}_1$ and $\mathcal{D}$. All the networks from the collection are trained separately on the respective erroneous states.}
    \label{fig:QAEs_erasure}
\end{figure}

\begin{figure}[t]
    \centering
    \includegraphics[width=0.99\linewidth]{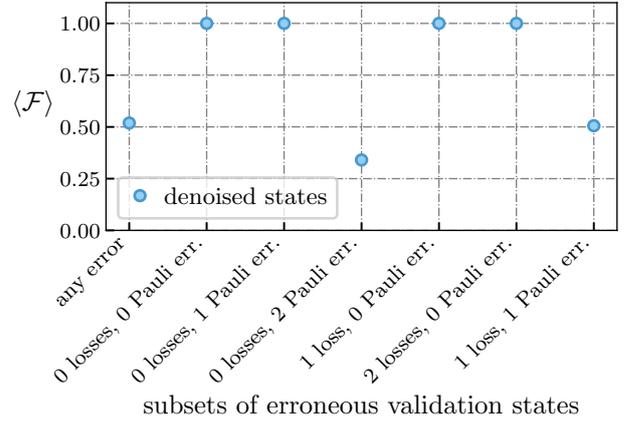}
    \caption{Validation of an $x$-1-5 collection of QAEs trained to denoise logical states of the 5-qubit code undergoing erasures with probability $p_{\mathrm{loss}} = 0.4$ followed by depolarizing noise occurring independently on every qubit with probability $p_{\mathrm{comp}} = 0.1$. The QAEs learn to correct arbitrary single-qubit Pauli errors and up to two arbitrary erasures of qubits. The collection of networks is tested on $5 \times 10^4$ randomly drawn validation states subjected to errors as outlined above.}
    \label{fig:515_results_loss}
\end{figure}

Here, we want to use QAEs to correct possible losses of physical qubits from logical quantum states.
We consider the quantum erasure channel, thus, the positions of losses are known.
Note that we restrict the investigation to erasures occurring on states before they enter a DQNN.
Since erasures of different qubits are classically distinguishable, it is possible to apply a different recovery map for the correction of every possible erasure event.
In this work we thus use a separate QAE for every possible loss.
Together, these QAEs form a collection of networks, as depicted in Fig.~\ref{fig:QAEs_erasure}.
The individual QAEs from the collection have to be trained separately on the respective erroneous states.
We model erasures by tracing over the corresponding qubits of a logical state, leaving behind a state that is generally mixed.
Networks that are used to correct single erasures on logical states of a code consisting of $n$ physical qubits thus implement channels mapping $(n-1)$-qubit states to $n$-qubit states.
In the following we will employ the self-inverse architecture to train the network that processes states which did not suffer any erasures.
The trained decoding channel $\mathcal{D}$ will then be reused for all other networks of the collection such that they differ only with regard to the encoding channels $\mathcal{E}_{i}$, as indicated in Fig.~\ref{fig:QAEs_erasure}.

As a minimal example we train a collection of $x$-1-4 QAEs on logical states of the 4-qubit erasure code being subjected to losses.
We find that after sufficient training the networks learn to perfectly correct any single erasure while failing to correct two or more losses.
In experimental devices for quantum information processing, erasures and computational errors occur side by side.
Thus, as a further example we train a collection of $x$-1-5 QAEs to correct errors on logical states of the 5-qubit error-correcting code.
The states suffer independent losses of qubits with probability $p_{\mathrm{loss}} = 0.4$ followed by depolarizing noise of strength $p_{\mathrm{comp}} = 0.1$.
Fig.~\ref{fig:515_results_loss} shows the performance of the $x$-1-5 QAEs to correct errors on randomly drawn logical states subjected to losses and Pauli errors, with the same probabilities as for the training process.
In particular, the plot shows denoising results for different subsets of errors.
We see that the QAEs learn to correct single Pauli errors as well as any single or double erasure.
This confirms our expectation regarding what is possible to achieve on logical states of the 5-qubit error-correcting code.


\section{Encoding Discovery}\label{section:encoding_discovery}

\begin{figure}
    \centering
    \includegraphics[width=\linewidth]{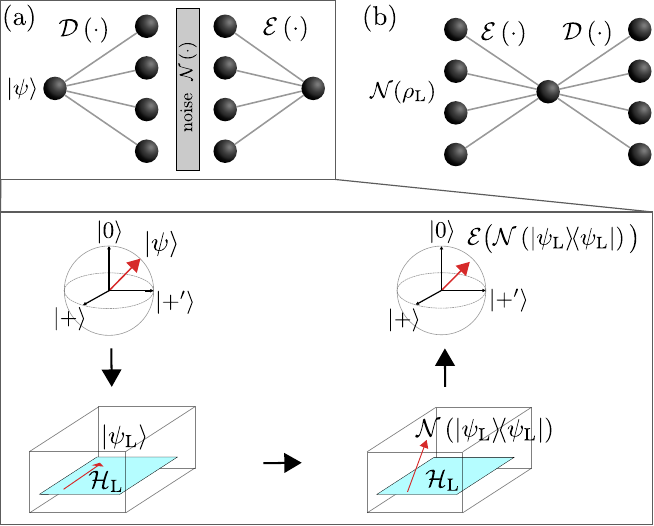}
    \caption{(a) Sketch of a quantum neural network that is used to unveil novel logical encodings protecting quantum information from noise $\mathcal{N}$. The network is trained to reproduce input states $\ket{\psi}$ at the output layer. The intermediate states $\mathcal{D}(\ketbra{\psi}) = \rho_{\mathrm{L}}$ correspond to a logical encoding of the input information that is learned in an unsupervised manner. Those states are subjected to noise $\mathcal{N}$ whereupon the channel $\mathcal{E}$ conducts a combined error correction and compression of the possibly faulty logical states. A good retrieval of the original single-qubit states will be possible if the network finds a logical encoding $\rho_{\mathrm{L}} = \ketbra{\psi_{\mathrm{L}}}$ that is well suited to deal with the present noise. (b) Interchanging the encoder and the decoder of the network in (a) gives rise to an undercomplete QAE that can be used to perform quantum error correction on erroneous logical states $\mathcal{N}(\rho_{\mathrm{L}})$.}
    \label{fig:EncodingFinder_Intro}
\end{figure}

In the previous section, we showed that quantum autoencoders are able to discover optimal strategies to correct computational errors and erasures on states from a predefined logical codespace.
However, the denoising capabilities of QAEs are fundamentally limited by the logical encoding defined in advance.
To allow for more flexible denoising strategies it would be desirable to search for entirely new logical encodings that are optimally suited to protect quantum information from unknown types of noise.
Some schemes have already been proposed on how quantum neural networks can be used to achieve this \cite{johnson2017,cong2019, cao2022quantumvariational}.
Here, we choose quantum neural networks in the spirit of overcomplete autoencoders as shown in Fig.~\ref{fig:EncodingFinder_Intro}(a) to solve this task.
The networks are trained to reconstruct the single-qubit input states at the output layer while noise is present in the interior of the network.
In a first step, a trainable channel $\mathcal{D}$ maps the single-qubit input states to unspecified logical states: $\rho_{\mathrm{L}} = \mathcal{D}(\ketbra{\psi})$.
The logical states are subjected to noise $\mathcal{N}$, corresponding to decoherence while the states are stored in memory.
The subsequent channel $\mathcal{E}$ is trained to map these states back to the original single-qubit inputs.
Thus, the network finds a suitable logical encoding that allows for errors induced by the intermediate noise channel to be corrected.
Such an optimal encoding is found in an unsupervised manner.
We propose then to rearrange the channels $\mathcal{D}$ and $\mathcal{E}$ of a trained network to form an undercomplete QAE with a single neuron on the central layer, as sketched in Fig.~\ref{fig:EncodingFinder_Intro}(b).
This QAE is ready to perform QEC on logical states defined by the newly discovered encoding rule given by $\mathcal{D}$, without the need to perform further training.
Exposing states $\rho_{\mathrm{L}}$ to noise and feeding those to the new QAE results in corrected logical states at the output layer.

To investigate whether the proposed QNNs can actually find suitable logical encodings and correction strategies we consider the following error model.
The qubits of a state are subjected to spatially correlated dephasing noise followed by possible erasures.
The collective dephasing arises from coherent $Z$-rotations of all qubits in the register, occurring probabilistically:
\begin{subequations}\label{eq:collective_dephasing}
    \begin{equation}
        \mathcal{N}^{\mathrm{coll.deph.}}(\rho) = \int p(\alpha) U(\alpha) \rho U(\alpha)^{\dagger} \mathrm{d}\alpha ,
    \end{equation}
    where
    \begin{equation}
        U(\alpha) = e^{-i \frac{\alpha}{2} \sum_n Z_n}
    \end{equation}
\end{subequations}
and the quantity $p(\alpha)$ describes a probability distribution.
Collective dephasing of idling qubits is a relevant type of noise in experimental setups of quantum processors.
For instance, it occurs in ion-trap devices because of global fluctuations of the magnetic field strengths \cite{monz2011}.
Collective dephasing noise can have detrimental effects on stored quantum information such as enhanced decoherence of entangled multi-qubit states \cite{monz2011,yu2003,bermudez2019}.
However, quantum information can be perfectly protected from collective dephasing noise by encoding logical states in a decoherence-free subspace (DFS) \cite{lidar1998,kwiat2000,carnio2015}.

Here, we train a collection of 1-4-1 QNNs, based on the sketch in Fig.~\ref{fig:EncodingFinder_Intro}(a), and expose the 4-qubit states in the center of the quantum networks to collective dephasing noise according to Eq.~\eqref{eq:collective_dephasing}, where $p(\alpha)$ is chosen to be a centered Gaussian with unit variance.
Moreover, erasures of single qubits may occur on the intermediate states, triggering one of the channels $\mathcal{E}_i$, $i = 0, \ldots, 4$ from the collection to perform the combined compression and correction of faulty states.
We employ the self-inverse architecture, meaning that the dissipative channel $\mathcal{E}_{0}$ embeds the inverse of the 5-qubit unitary matrix realizing the map $\mathcal{D}$.
We find that the collection of quantum networks learns to encode logical states in a DFS.
Moreover, the encoding allows for single erasures of qubits to be corrected.
Logical states are thus perfectly protected from collective dephasing and partially protected from losses of qubits.
In Appendix~\ref{appendix:encoding} we show the precise form of the discovered logical states and discuss the numerical results in more detail.


\section{Robustness Against Internal Noise}\label{section:internal_noise}

So far, we have assumed that noise only acts on incoming qubits while the DQNNs themselves operate perfectly.
This is, however, an idealization.
In general, the application of gates as well as idling of qubits during a computation will introduce errors on the quantum states operated.
In this work we focus on a quantum memory setting, i.e.~stored quantum information to be protected from noise.
One goal is to extend the lifetime of a logical qubit beyond the lifetime of a bare physical qubit.
Reaching this ``break-even'' point is a present-day challenge.
Using QEC to extend the lifetime of an encoded qubit has been demonstrated experimentally with bosonic codes \cite{ofek2016,hu2019}.
In other works, active quantum error correction was shown to be advantageous in some specific noise parameter regimes~\cite{schindler2011}.

\begin{figure}
    \centering
    \includegraphics[width=\linewidth]{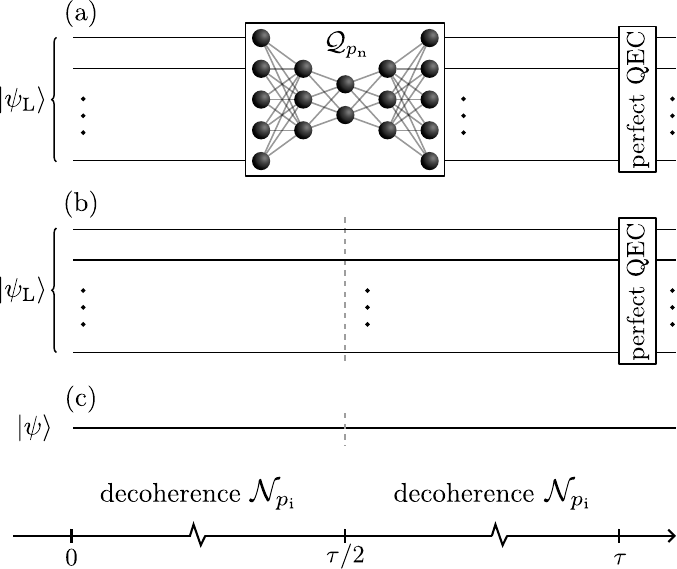}
    \caption{Scheme to evaluate the efficacy of QEC with a QAE in a quantum memory. A quantum state, either an encoded logical state $\psiL$ (a,b) or a single-qubit state $\ket{\psi}$ stored in a bare physical qubit (c), is stored in memory for a time $\tau/2$ such that noise $\mathcal{N}_{\pidle}$ acts on all physical qubits. We may now choose to perform a round of imperfect QEC (a), or to not apply it (b). Here, we assume, for simplicity, that the QEC round happens on a much shorter timescale than $\tau$. The qubits then idle for another time $\tau/2$ which introduces noise $\mathcal{N}_{\pidle}$. Finally, logical states are projected back to the codespace by a round of perfect QEC, which provides a tool to assess the state quality. The fidelity of the output w.r.t.~the initial state determines the probability $\mathcal{P}$ of successful state discrimination and serves as a measure for the quality of the quantum memory.}
    \label{fig:noisy_QEC_setting}
\end{figure}

To assess whether we can use an intrinsically noisy QAE to extend the lifetime of an encoded qubit, we employ a measure proposed and discussed in Ref.~\cite{bermudez2017}, as sketched in Fig.~\ref{fig:noisy_QEC_setting}.
The goal is to protect quantum information from environmental decoherence for a time interval $\tau$.
To do so, one can either use a quantum state $\ket{\psi}$ on a bare physical qubit or decide to encode the information into a logical multi-qubit state $\psiL$.
In any case, all physical qubits are subjected to noise while being stored in memory.
Now, the question arises whether it is beneficial to apply a round of imperfect quantum error correction to the encoded state after, say, half of the memory time, to correct errors that have accumulated thus far.
It is not surprising that a very noisy QEC device rather deteriorates the encoded state than improving it.
A ``good'' QEC device, however, can actually yield an advantage, as compared to doing nothing.
To assess the usefulness of a noisy QEC device we therefore compare three scenarios.
One starts either with a single-qubit quantum state $\ket{\psi}$ or an encoded logical state $\psiL$.
All qubits are subjected to noise $\mathcal{N}_{\pidle}$ caused by environmental decoherence while being stored in memory for a time $\tau/2$.
Now, one can apply a round of imperfect QEC to the encoded state, assuming, for simplicity, that the application of the QEC cycle happens on a much shorter timescale than the idling time $\tau$.
Afterwards, the states are left in memory for a further time $\tau/2$, again inducing noise $\mathcal{N}_{\pidle}$.
Finally, we project encoded logical states back to the codespace by performing a round of perfect QEC.
We note that this last step is not part of an actual experimental protocol but rather serves as a tool for the quantitative assessment of the state's quality.
The probability $\mathcal{P}$ of successful state discrimination is then given by the fidelity of a final state w.r.t. the corresponding noise-free initial state.
We take this quantity as a measure for the quality of the quantum memory.
Comparing the values of $\mathcal{P}$ for the different scenarios in Fig.~\ref{fig:noisy_QEC_setting} informs us whether the application of the intrinsically noisy QEC device can be beneficial.

\begin{figure}[t]
    \centering
    \includegraphics[width=\linewidth]{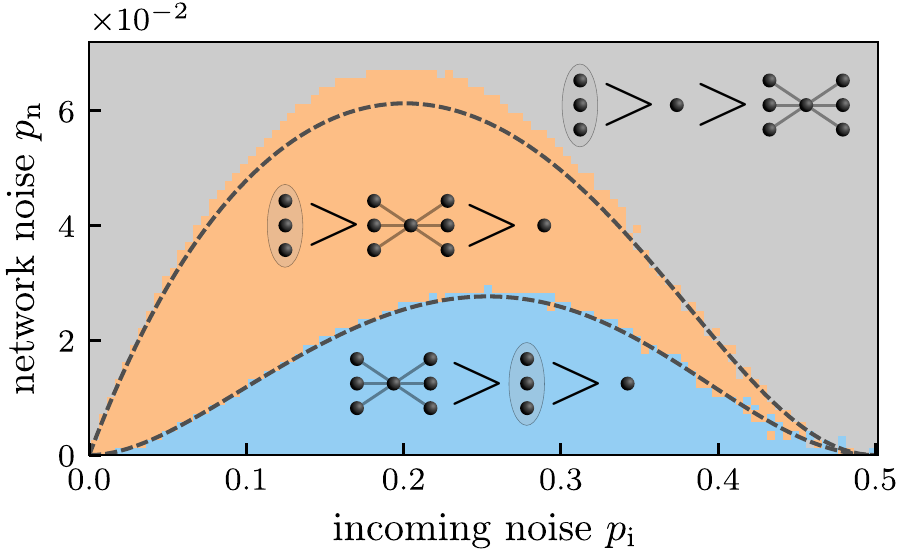}
    \caption{Phase diagram indicating the benefit of a QAE for QEC. It shows for which values of incoming bit flip noise strengths $\pidle$ and network noise strengths $\pnetwork$ a 3-1-3 QAE is beneficial for active QEC in a quantum memory. The range of noise strengths in which the QAE-corrected memory is superior to both a bare physical qubit and an encoded but uncorrected qubit is shown as the lower blue region. The orange-colored intermediate region corresponds to the actively corrected qubit performing better than a bare qubit but worse than the uncorrected logical qubit. In the grey-colored upper region, applying the QAE for QEC is inferior to both other cases. Each data point is obtained by randomly drawing $10^4$ logical states, exposing them to a bit flip channel and processing these states $\mathcal{N}_{\pidle}(\ketbra{\psi_{\mathrm{L}}})$ with the noisy QAE. Then, another round of noise $\mathcal{N}_{\pidle}$ and a perfect round of QEC are applied. We compare the probability of successful state discrimination to that of a physical qubit and an uncorrected encoded qubit. The phase boundaries are in some parts pixelated due to finite sampling statistics. The lines correspond to phase boundaries predicted from an analytical expansion to first order in $\pnetwork$ (see Appendix~\ref{appendix:noisy_QAE}). For small values of $\pnetwork$ the analytical boundaries match the actual boundaries very well, whereas small deviations for larger values of $\pnetwork$ are expected.}
    \label{fig:phase_diagram}
\end{figure}

To investigate whether noisy QAEs introduced in this paper can prove beneficial for quantum error correction, we analyze a minimal example, which is an intrinsically noisy \mbox{3-1-3} QAE to correct bit flip errors.
We note, however, that the analysis we perform for this example could also be applied to larger codes and therefore other types of network structures and noise.
In this work we do not focus on a specific physical platform and therefore consider a platform-agnostic noise model.
In particular, we apply a multi-qubit depolarizing channel after every application of a unitary matrix.
In this channel, any-weight Pauli errors occur with equal probabilities $\pnetwork/(4^m - 1)$, where $m$ is the number of qubits that the noise channel acts upon.
Here we consider a \mbox{3-1-3} DQNN with standard architecture consisting of four independent unitary matrices, i.e.~we do not use the self-inverse ansatz.
A corresponding quantum circuit representation can be found in Appendix~\ref{appendix:noisy_QAE}.
The internal noise thus acts at four positions in the circuit.
We model the environmental decoherence in the quantum memory as bit flip errors occurring independently on every qubit with probability $\pidle$.
\begin{figure}[t]
    \centering
    \includegraphics[width=0.99\linewidth]{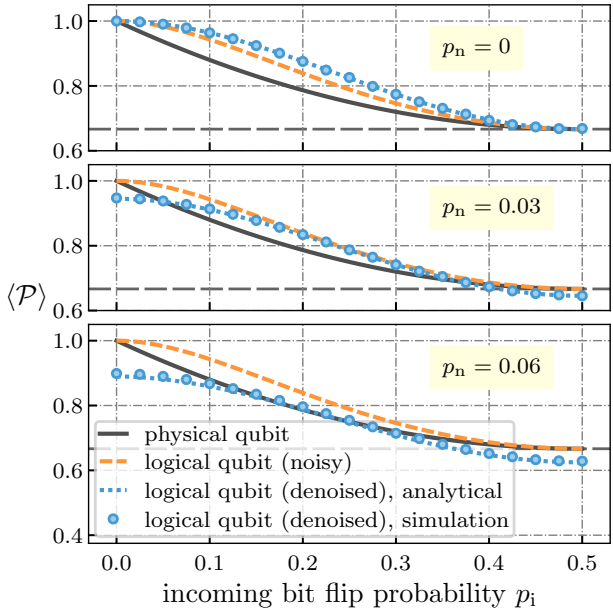}
    \caption{Comparison of the three different quantum memory settings shown in Fig.~\ref{fig:noisy_QEC_setting}. A bare physical qubit (solid black line), an encoded but uncorrected logical qubit (dashed orange line) and a corrected logical qubit (dotted blue line and data points) are compared. As incoming noise we consider independent bit flips occurring with probability $\pidle$ on every qubit. We use logical states of the three-qubit repetition code and apply a 3-1-3 QAE for QEC. The QAE is intrinsically noisy with the strength of the network noise quantified by $\pnetwork$. Encoding quantum information in logical states and performing active QEC is advantageous if the probability of successful state discrimination $\mathcal{P}$ of denoised states is larger than for single-qubit and uncorrected logical states. A feature of the three-qubit code is that the quality of the latter two saturates at $\langle \mathcal{P} \rangle = 2/3$ for $\pidle = 0.5$, indicated by the horizontal dashed line. For a perfect QAE, applying active QEC is beneficial for $\pidle < 0.5$. For an increasingly noisy QAE, the range of incoming noise strengths for which it is beneficial to apply QEC is reduced. The numerical data is obtained by averaging over $10^4$ randomly drawn logical states for each data point.}
    \label{fig:noisy_QEC}
\end{figure}
For various pairs of noise strengths $(\pidle, \pnetwork)$ we perform numerical simulations according to Fig.~\ref{fig:noisy_QEC_setting} to analyze how well the intrinsically noisy QAE corrects bit flip errors on the logical states.
Fig.~\ref{fig:noisy_QEC} compares the quality of the three quantum memory settings introduced in Fig.~\ref{fig:noisy_QEC_setting} for various different network noise strengths $\pnetwork$ and incoming noise strengths $\pidle$.
We average the probability of successful state discrimination over a large number of different input states.
First, we note that for bit flip probabilities $\pidle < 0.5$ the uncorrected encoded qubit performs better than the bare physical qubit. This is a known property of the repetition code, resulting from the correctability of all single-qubit errors in the final perfect round of QEC.
For a noise-free QAE, $\pnetwork = 0$, the error correction is advantageous for incoming noise strengths in exactly that range.
Small values of $\pnetwork$ reduce the interval of incoming noise strengths for which performing QEC is still beneficial.
Above a certain threshold of $\pnetwork$, the quality of an actively corrected logical qubit drops below the quality of an uncorrected encoded qubit.
If we increase $\pnetwork$ even further, the scheme involving the noisy QAE is eventually outperformed by a bare physical qubit.
In Appendix~\ref{appendix:noisy_QAE} we derive an approximate expression for the quality of denoised logical states as a function of $\pidle$ and $\pnetwork$.
We find that it scales linearly with $\pnetwork$, highlighting that the design of the QAE is not fault-tolerant.
This is, however, not surprising, since errors on the single bottleneck qubit inevitably result in logical errors on the final state.
Performing an extensive analysis for various pairs of noise strengths $(\pidle, \pnetwork)$ we obtain a phase diagram, depicted in Fig.~\ref{fig:phase_diagram}, that indicates three regimes: one where the application of a noisy QAE for quantum error correction is advantageous compared to both a bare physical qubit and an encoded but uncorrected logical qubit; a regime where the quality of the actively corrected qubit lies between the latter two; and a third regime where the actively corrected qubit is outperformed by the other two approaches.
We see that noisy QAEs can be successfully used for QEC as long as the internal noise of the quantum network stays below a certain $\pidle$-dependent threshold.
To obtain an approximate form of the phase boundaries we expand the output states of the noisy 3-1-3 QAE up to linear order in $\pnetwork$.
The calculation is sketched in Appendix~\ref{appendix:noisy_QAE}.
For small values of $\pnetwork$ we observe excellent agreement of the analytical phase boundaries with the numerical results.
For encoded logical states in the quantum memory, the final round of perfect QEC removes any single bit flip errors, such that the logical error rate and thus also the probability of successful state discrimination are quadratic in $\pidle$.
Therefore, also the boundary separating the blue-colored and orange-colored regions in Fig.~\ref{fig:phase_diagram} shows a quadratic behavior for small values of $\pidle$.
In contrast to this finding, the boundary which separates the orange-colored and the grey-colored regions is linear for small $\pidle$.
This results from the fact that the probability of successful state discrimination for a single-qubit state is linear in $\pidle$.
Typically, we are interested in the regime of small, though not too small incoming noise and small network noise strengths.
In that case the phase boundary separating the blue and the orange phase can be very well approximated as $\pnetwork^{\mathrm{crit., \, logical}} = \frac{765}{351} \pidle^2$, as derived in Appendix~\ref{appendix:noisy_QAE}.
For the investigated QAE consisting of four unitary matrices the probability for a single error to occur during the application of the network is in leading order $p_{\mathrm{1 \, err.}} \approx 4 \pnetwork$.
This means that for $p_{\mathrm{1 \, err.}} \lesssim 4\cdot \frac{765}{351} \pidle^2 = \frac{3060}{351} \pidle^2$ it is for this network structure advantageous to apply a QAE for error correction in the quantum memory.


\section{Discussion and Outlook}\label{section:conclusion}

In this paper we showed that quantum neural networks in the form of quantum autoencoders can be used to perform quantum error correction.
QAEs are able to correct errors on arbitrary states from a predefined logical codespace such that the lifetime of encoded logical qubits can be enhanced.
In particular, if a logical encoding allows for various error correction strategies to be applied, a QAE can learn the strategy yielding the best possible denoising results.
As an example we demonstrated that QAEs are able to adapt to spatially correlated bit flip noise.
Moreover, we showed for the first time that the error correction abilities of QAEs are not limited to computational errors but extend also to the correction of qubit erasures.
Other types of QNNs designed in the spirit of overcomplete quantum autoencoders can be used to find novel logical encodings being optimally suited to correct hardware-specific noise.
We proposed and showed that these networks can be directly transformed into undercomplete QAEs ready to perform QEC on the discovered logical states without the need to perform further training.
Lastly, even QAEs that are intrinsically noisy can be used successfully for QEC in a quantum memory if the internal noise of the quantum neural network is sufficiently low.

However, we note that we encountered difficulties in the training of DQNNs for QEC. We observe that especially the training of encoding finder networks, as discussed in Sec.~\ref{section:encoding_discovery}, frequently converges towards solutions yielding non-optimal error correction strategies.
These observations indicate the existence of saddle points or local minima in the cost function landscape.
At the beginning of the training process, the unitary matrices composing a DQNN are initialized randomly, so convergence towards those non-optimal points is hard to avoid using standard gradient descent methods.
To obtain a quantum network that is able to optimally correct errors, we thus have to perform a repeated random initialization of the training parameters followed by a gradient descent algorithm.
These findings are, however, not unexpected since the occurrence of barren plateaus, where cost function gradients vanish exponentially in the number of qubits \cite{mcClean2018}, is frequently observed for variational quantum algorithms.
These training issues therefore impede a straightforward scaling to substantially larger quantum neural networks, where we expect the training difficulties to become more prominent for increasing depth and width of the networks.
In fact, it has recently been shown explicitly that also DQNNs are affected by barren plateaus \cite{sharma2022}.
Ongoing research on the trainability of QNNs involves e.g.~the investigation of different cost functions such that the occurrence of barren plateaus can be avoided \cite{cerezo2021cost}.
Moreover, it is a challenge to find suitable variational ans{\"a}tze that are sufficiently expressive while avoiding the occurrence of barren plateaus \cite{holmes2022}.
Eventually, one aims at finding hardware-specific ans{\"a}tze with a reduced number of parameters that allow for an easy application of a QNN on available quantum hardware and do not suffer from serious trainability problems.

In summary, our work shows that QAEs could serve as a versatile tool for autonomous quantum error correction of a wide variety of error sources and characteristics in a quantum memory.
The QAE framework is especially attractive for experimental setups where in-sequence measurements with real-time feedback as required for the conventional QEC approach are not readily available or inefficient.
Thus, a comparison of quantum autoencoders with other autonomous QEC proposals would be interesting.
Furthermore, possible future work could include the investigation of fault-tolerant designs of QAEs for quantum error correction to extend their applicability beyond the quantum memory setting.
In this context, one could also imagine logical qubits corrected by QAEs being used as low-level autonomously running units that are trained to deal with the dominant error sources of a given architecture.
Those could then form  building blocks of more complex established QEC codes, in analogy to concatenating basic few-qubit codes with more advanced codes \cite{cross2009} or using bosonic codes as building blocks for scalable QEC schemes \cite{terhal2020}.


\begin{acknowledgments}
We thank D. Bondarenko, P. Feldmann and D. DiVincenzo for stimulating discussions and M. Rispler for feedbacks on the manuscript.
We acknowledge support by the ERC Starting Grant QNets Grant Number 804247,
the EU H2020-FETFLAG-2018-03 under Grant Agreement number 820495,
by the German ministry of science and education (BMBF) via the VDI within the project IQuAn,
by the Deutsche Forschungsgemeinschaft through Grant No. 449905436,
and by US A.R.O. through Grant No. W911NF-21-1-0007,
and by the Office of the Director of National Intelligence (ODNI), Intelligence Advanced Research Projects Activity (IARPA), via US ARO Grant number W911NF-16-1-0070.
All statements of fact, opinions or conclusions contained herein are those of the authors and should not be construed as representing the official views or policies of ODNI, the IARPA, or the US Government.
The network coding and training was done in the Matlab programming language, based on code available on Github at~\cite{beer2020,bondarenko2020}.
\end{acknowledgments}

\bibliographystyle{quantum}
\bibliography{references}


\appendix
\newpage
\onecolumngrid

\section{Quantum Process Tomography}\label{appendix:QPT}

In this appendix we study some of the quantum autoencoders discussed in the main text in more detail by analyzing the maps which they implement using quantum process tomographies.
The error correction map of the standard 3-qubit bit flip code is given by
\begin{subequations}\label{eq:3QC_Krausmap}
	\begin{equation}
		\mathcal{R}(\cdot) = \sum_{b}^{} M_{b} \, \cdot \, M_{b}^{\dagger} \, ,
	\end{equation}
	with Kraus operators
	\begin{equation}
		\begin{aligned}
			M_{00} & = \ketbra{000}{000} + \ketbra{111}{111} \\
			M_{01} & = \ketbra{000}{001} + \ketbra{111}{110} \\
			M_{10} & = \ketbra{000}{100} + \ketbra{111}{011} \\
			M_{11} & = \ketbra{000}{010} + \ketbra{111}{101} .
		\end{aligned}
	\end{equation}
\end{subequations}
In Sec.~\ref{section:QEC_results} of the main text we demonstrate that our fully trained 3-1-3 QAEs correct bit flip errors on 3QC logical states as well as the standard 3QC does.
To prove the assumption that the QAEs implement the map stated in Eq.~\eqref{eq:3QC_Krausmap}, we perform quantum process tomographies of the channels realized by the networks.
Given an operator basis $\{E_i\}$, a quantum channel $\mathcal{Q}$ can be written as
\begin{equation}\label{}
	\mathcal{Q}(\cdot) = \sum_{i,j} \chi_{ij} E_{i} \cdot E_{j}^{\dagger} ,
\end{equation}
where the complex Hermitian matrix $\chi$ uniquely characterizes the quantum channel $\mathcal{Q}$ w.r.t.~the basis $\{E_i\}$ \cite{nielsenChuang2010}.
Figure~\ref{fig:QPT_313}(c) in the main text shows the process matrix $\chi$ of a fully trained 3-1-3 QAE which is equal to the process matrix of the map in Eq.~\eqref{eq:3QC_Krausmap}.
Analyzing the trained QAEs shown in Fig.~\ref{fig:313_results} in the main text, we find that all process matrices apart from the one corresponding to the network trained on $p = 0$ equal the matrix in Fig.~\ref{fig:QPT_313}(c).
This demonstrates that QAEs which are trained on noisy states learn to implement the correction channel of the 3-qubit bit flip code, even though the number of different training states is very limited.

The analytical curve describing the performance of the standard 3-qubit code in Fig.~\ref{fig:313_results} can be obtained as follows.
As stated in the main text, a logical 3QC state $\psiL$ undergoing independent bit flip noise of strength $p$ and being actively corrected afterwards suffers a logical error with probability $\pL = 3p^2(1-p) + p^3$ and no error with probability $1 - \pL$, thus resulting in a state
\begin{equation}
    \rho_{\mathrm{denoised}} = (1-\pL) \ketbra{\psi_{\mathrm{L}}} + \pL \XL \ketbra{\psi_{\mathrm{L}}} \XL .
\end{equation}
For pure states $\psiL = \cos(\theta/2) \zeroL + e^{i \phi} \sin(\theta/2) \oneL$, the averaged fidelity of denoised states w.r.t.~the noise-free target states therefore reads
\begin{equation}
    \langle \mathcal{F} \rangle = \iint \frac{\mathrm{d}\theta \, \mathrm{d}\phi}{4 \pi} \sin(\theta) \bra{\psi_{\mathrm{L}}} \rho_{\mathrm{denoised}} \ket{\psi_{\mathrm{L}}} = 1 - \frac{2}{3} \pL .
\end{equation}

\begin{figure}[t]
    \centering
    \includegraphics[width=1.0\linewidth]{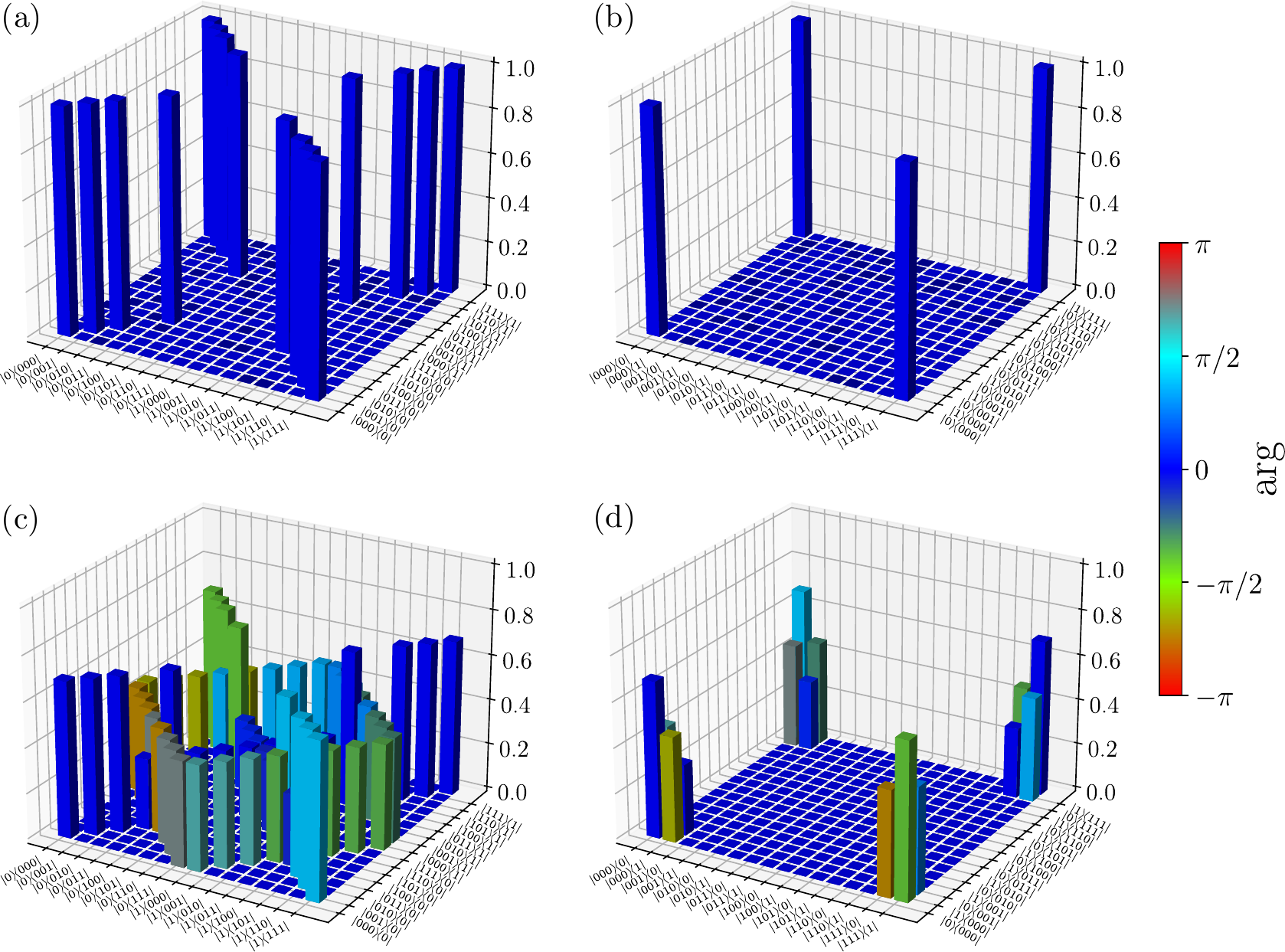}
    \caption{(a) Quantum process tomography of the quantum channel described by the Kraus operators in Eq.~\eqref{eq:31_encoding} which simultaneously corrects single bit flips on logical states of the 3QC and compresses them to single-qubit states. (b) Process matrix of the channel with Kraus operator in Eq.~\eqref{eq:13_decoding} which reconstructs logical states of the 3QC. (c) Process matrix of the 3-1 encoding channel of a trained QAE. (d) Quantum process tomography of the 1-3 decoding channel of a trained QAE. Performing the maps in (c) and (d) successively results in the same map that is obtained when the channels in (a) and (b) are applied one after another.}
    \label{fig:QPT_encoder_decoder}
\end{figure}

Another point which we discuss in the main text is that the encoding channel of a 3-1-3 QAE conducts the combined compression and correction of erroneous input states while the decoding channel performs a trivial reconstruction of logical states.
The single-qubit intermediate state can, however, be arbitrarily rotated against the logical input/output state.
This can be seen from quantum process tomographies of the individual encoding and decoding channels for various trained 3-1-3 QAEs.
Fig.~\ref{fig:QPT_encoder_decoder}(a) shows the process matrix of a handmade encoding channel described by the Kraus operators
\begin{equation}\label{eq:31_encoding}
	\begin{aligned}
		M_{00} & = \ketbra{0}{000} + \ketbra{1}{111} \\
		M_{01} & = \ketbra{0}{001} + \ketbra{1}{110} \\
		M_{10} & = \ketbra{0}{100} + \ketbra{1}{011} \\
		M_{11} & = \ketbra{0}{010} + \ketbra{1}{101} .
	\end{aligned}
\end{equation}
The quantum process tomography of the corresponding decoding channel, characterized by a single Kraus operator
\begin{equation}\label{eq:13_decoding}
    M_{0} = \ketbra{000}{0} + \ketbra{111}{1} ,
\end{equation}
is shown in Fig.~\ref{fig:QPT_encoder_decoder}(b).
For an exemplary 3-1-3 QAE trained to implement the 3QC error correction map, Figs.~\ref{fig:QPT_encoder_decoder}(c) and (d) show process matrices of the respective encoding and decoding channels.
The concatenation of the two maps yields the channel given by Eq.~\eqref{eq:3QC_Krausmap}.
Compared to the handmade channels we see, however, that the single-qubit intermediate state is rotated against the logical input and output state.


\section{Training QAEs with Self-inverse Architecture}\label{appendix:training_backpropagation}

Beer \textit{et al.} \cite{beer2020} describe how DQNNs can be trained efficiently in numerical simulations.
At time step $s$ the gradient descent step is performed by updating all unitary matrices in the network according to the rule
\begin{equation}
    U_{k}^{j_{k}}(s+\epsilon) = e^{i \epsilon K_{k}^{j_{k}}(s)} U_{k}^{j_{k}}(s) ,
\end{equation}
where the update matrices $K_{k}^{j_{k}}$ are chosen such that the cost function
\begin{equation}\label{}
	C(s) = 1 - \frac{1}{N} \sum_{i=1}^{N} \bra{\phi_{\mathrm{targ}}^{i}} \rho_{\mathrm{out}}^{i}(s) \ket{\phi_{\mathrm{targ}}^{i}}
\end{equation}
is reduced as fast as possible.
A careful analysis yields a simple rule to calculate the update matrices $K_{k}^{j_{k}}$, resembling a backpropagation algorithm.
The training input states $\rhoin^i$ are propagated forward through the network while the corresponding target states $\ket{\phi_{\mathrm{targ}}^{i}}$ are propagated backwards.
Commutators between the density matrices of the forward- and backpropagted states yield the quantities
\begin{equation}\label{eq:commutator_forward_backward}
	\begin{aligned}
		M_{k}^{j_k}(s,i) & = \Big[ U_{k}^{j_k}(s) \ldots U_2^1(s) \Big( \rho_{\mathrm{in}}^{i} \otimes \ketbra{0 \ldots 0}_{\mathrm{in, hidden}} \Big) \big(U_2^{1}(s)\big)^{\dagger} \ldots \big(U_{k}^{j_k}(s)\big)^{\dagger} \ , \\
		& \big(U_{k}^{j_k + 1}(s)\big)^{\dagger} \ldots \big(U_{\mathrm{out}}^{j_{\mathrm{out}}^{\mathrm{max}}}(s)\big)^{\dagger} \Big( \mathds{1}_{\mathrm{in, hidden}} \otimes \ketbra{\phi_{\mathrm{targ}}^{i}} \Big) U_{\mathrm{out}}^{j_{\mathrm{out}}^{\mathrm{max}}}(s) \ldots U_{k}^{j_k + 1}(s) \Big] ,
	\end{aligned}
\end{equation}
which are involved in the calculation of the update matrices:
\begin{equation}\label{}
	K_{k}^{j_k}(s) = i \frac{\mathrm{dim}(U_{k}^{j_k})}{2N} \sum_{i=1}^{N} \underset{\begin{array}{c} \mathrm{\scriptstyle qubits} \\[-4pt] \mathrm{\scriptstyle not \ in} \ \scriptstyle U_{k}^{j_k} \end{array}}{\Tr} \! \! \! \! \Big[ M_{k}^{j_k}(s,i) \Big] .
\end{equation}
For a detailed description the reader is referred to Ref.~\cite{beer2020}.

\begin{figure}
    \centering
    \includegraphics[width=.55\linewidth]{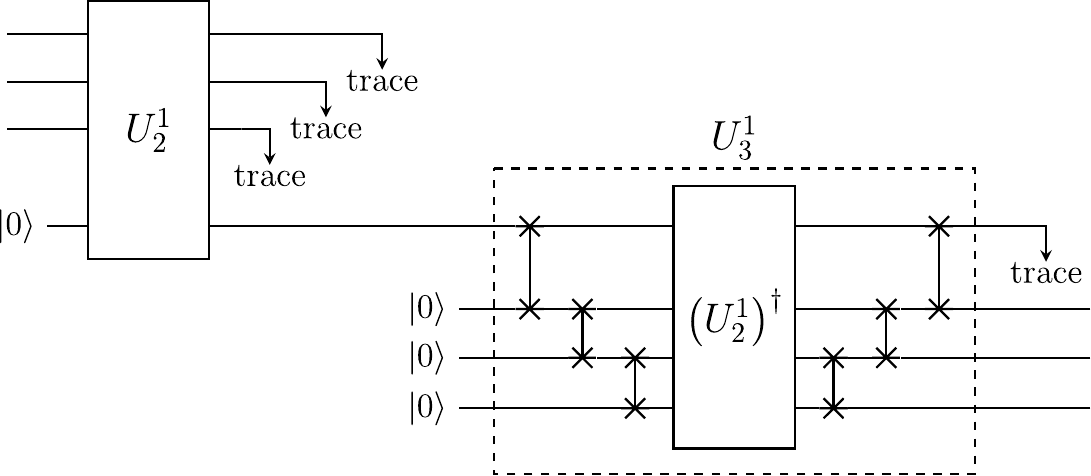}
    \caption{Quantum circuit realizing a 3-1-3 QAE with self-inverse architecture. The inverse of the matrix $U_2^1$ which occurs in the encoding channel is employed in the decoding channel. Since the three qubits belonging to the output layer are added after the intermediate qubit, a set of swap gates is required to permute the qubit indices.}
    \label{fig:circuit_recurrent_QAE_2}
\end{figure}

However, for QAEs with self-inverse architecture used in this paper the training algorithm requires some modifications which we describe in the following.
These adjustments arise from the fact that the same unitary matrices occur at several positions in the network.
For a numerical implementation it is convenient to always introduce new qubits of succeeding layers at the tail end of the latest state.
Therefore, the unitary matrices used for the decoding channel are not just the inverses of the previously used matrices but they come with a set of swap gates, as depicted in Fig.~\ref{fig:circuit_recurrent_QAE_2}.
However, these gates only perform a permutation of qubit indices, so they do not have to be applied physically in the network.
In a QAE with self-inverse architecture, a unitary matrix $U_{k}^{j_k}$ appearing in the encoding channel occurs in the decoder as the $\bar{j}_{k}$-th matrix realizing the transition to layer $\bar{k}$.
Abbreviating the necessary swap gates as $S_{k}^{j_k}$ yields
\begin{equation}\label{}
	U_{\bar{k}}^{\bar{j}_{k}} = S_{k}^{j_k} \big(U_{k}^{j_k}\big)^{\dagger} \big(S_{k}^{j_k}\big)^{\dagger} .
\end{equation}
We consider QAEs where the unitaries of the encoder act on all qubits of the preceding layer and a single qubit in the succeeding layer.
For a QAE consisting of $L$ layers (including the input layer), one finds $\bar{k} = L + 2 - k$ and $\bar{j}_{k} = n_{k} + 1 - j_k$, where $n_k$ is the width of the $k$-th layer.
Identical matrices occurring at several positions within the DQNN require the training update rules to be modified.
The goal is to find update matrices $K_{k}^{j_k}$ such that the unitary matrices from the encoder are again updated according to the rule
\begin{equation}\label{}
	U_{k}^{j_k}(s + \epsilon) = e^{i \epsilon K_{k}^{j_k}(s)} \, U_{k}^{j_k}(s) .
\end{equation}
The updates of the unitary matrices assembling the decoder, however, follow from the updated encoding matrices:
\begin{equation}\label{}
	\begin{aligned}
		U_{\bar{k}}^{\bar{j}_{k}}(s + \epsilon) & = S_{k}^{j_k} \left( U_{k}^{j_k}(s + \epsilon) \right)^{\dagger} \big(S_{k}^{j_k}\big)^{\dagger} \\
		& = S_{k}^{j_k} \left( \big(U_{k}^{j_k}(s)\big)^{\dagger} e^{-i \epsilon K_{k}^{j_k}(s)} \right) \big(S_{k}^{j_k}\big)^{\dagger} \\
		& = \left( S_{k}^{j_k} \big(U_{k}^{j_k}(s)\big)^{\dagger} e^{-i \epsilon K_{k}^{j_k}(s)} \, U_{k}^{j_k}(s) \big(S_{k}^{j_k}\big)^{\dagger} \right) U_{\bar{k}}^{\bar{j}_{k}}(s) .
	\end{aligned}
\end{equation}
We now derive how the update matrices $K_{k}^{j_k}$ are calculated such that the cost function $C$ is reduced as quickly as possible.
An expression for $\mathrm{d}C/\mathrm{d}s$ can be obtained:
\begin{equation}\label{}
	\begin{aligned}
		\frac{\mathrm{d} C(s)}{\mathrm{d} s} = \frac{-i}{N} \sum_{i=1}^{N} \Tr & \Big[ M_2^1(s,i) K_2^1(s) + \ldots + M_{(L+1)/2}^{j^{\mathrm{max}}}(s,i) K_{(L+1)/2}^{j^{\mathrm{max}}}(s) \\
		& + M_{(L+3)/2}^{1}(s,i) J_{(L+3)/2}^{1}(s) + \ldots + M_{\mathrm{out}}^{j^{\mathrm{max}}}(s,i) J_{\mathrm{out}}^{j^{\mathrm{max}}}(s) \Big] .
	\end{aligned}
\end{equation}
Here, $M_{k}^{j_k}$ is defined in Eq.~\eqref{eq:commutator_forward_backward} and the quantities $J_{\bar{k}}^{\bar{j}_{k}}$ are given by
\begin{equation}\label{}
	J_{\bar{k}}^{\bar{j}_{k}} = - S_{k}^{j_k} \big(U_{k}^{j_k}\big)^{\dagger} K_{k}^{j_k} \, U_{k}^{j_k} \big(S_{k}^{j_k}\big)^{\dagger} .
\end{equation}
A matrix $K_{k}^{j_k}$ thus occurs twice in the expression for $\mathrm{d}C/\mathrm{d}s$.
Minimizing it therefore yields additional terms in the expressions for $K_{k}^{j_k}$:
\begin{equation}\label{}
		K_{k}^{j_k}(s) = i \frac{\mathrm{dim}(U_{k}^{j_k})}{2N} \sum_{i=1}^{N} \underset{\begin{array}{c} \mathrm{\scriptstyle qubits} \\[-4pt] \mathrm{\scriptstyle not \ in} \ \scriptstyle U_{k}^{j_k} \end{array}}{\Tr} \! \! \! \! \! \! \Big[ M_{k}^{j_k}(s,i) \Big] \; \; - \! \! \! \! \! \underset{\begin{array}{c} \mathrm{\scriptstyle qubits} \\[-4pt] \mathrm{\scriptstyle not \ in} \ \scriptstyle U_{\bar{k}}^{\bar{j}_{k}} \end{array}}{\Tr} \! \! \! \! \! \! \Big[ U_{k}^{j_k}(s) \big(S_{k}^{j_k}\big)^{\dagger}  M_{\bar{k}}^{\bar{j}_{k}}(s,i) \, S_{k}^{j_k} \big(U_{k}^{j_k}(s)\big)^{\dagger} \Big] .
\end{equation}
An update matrix $K_{k}^{j_k}$ can therefore still be obtained by forward- and backpropagation of the training states.
However, since the corresponding unitary matrix $U_{k}^{j_k}$ occurs at two positions in the network, it is necessary to propagate the training states to both places, for calculating the respective commutators between forward- and backpropagated states and constructing the update matrix for that specific unitary.


\section{Training Specifications}\label{appendix:training_specifications}

Here we give details on the training of the DQNNs discussed in the main text.
All networks were trained numerically using the training algorithm described in Appendix~\ref{appendix:training_backpropagation}.
A good overview of gradient descent variants for the training of classical neural networks can be found in a review by Ruder \cite{ruder2017}.
The set of states being used for the training of a DQNN is called training batch.
We divide the batch into minibatches of size $S_{\mathrm{minibatch}}$ and perform a gradient descent step after states from one minibatch were exposed to the network.
Presenting all minibatches to the network is called a training epoch.
After a training epoch we shuffle all states in the batch, create new minibatches and continue the training for a total number of $N_{\mathrm{epochs}}$ training epochs.
The gradient descent algorithm uses a learning rate $\epsilon$.
Moreover, we use the Nadam gradient descent optimizer with memory coefficients $\beta_1 = 0.9$ and $\beta_2 = 0.999$ to achieve better training convergence \cite{bondarenko2020,ruder2017}.
Table~\ref{tab:hyperparameters} summarizes the hyperparameters that were used for the training of the QAEs discussed in the paper.
\begin{table}[H]
    \centering
    \begin{tabular}{l|l|l|l|l} 
        & $\epsilon$ & $N_{\mathrm{epochs}}$ & $S_{\mathrm{minibatch}}$ & training batch \\ [0.5ex] 
        \hline
        Fig.~\ref{fig:313_results} \ & 0.1 \ & 200 \ & 3 \ & $\ket{0_{\mathrm{L}}}, \, \ket{1_{\mathrm{L}}}, \, \ket{+_{\mathrm{L}}}$ \\
        Fig.~\ref{fig:515_results} & 0.2 & 200 & 2 & $\ket{0_{\mathrm{L}}}, \, \ket{1_{\mathrm{L}}}, \, \ket{+_{\mathrm{L}}}, \, \ket{-_{\mathrm{L}}}, \, \ket{+'_{\mathrm{L}}}, \, \ket{-'_{\mathrm{L}}}$ \\
        Fig.~\ref{fig:313_results_correlated} & 0.1 & 200 & 3 & $\ket{0_{\mathrm{L}}}, \, \ket{1_{\mathrm{L}}}, \, \ket{+_{\mathrm{L}}}, \, \ket{-_{\mathrm{L}}}, \, \ket{+'_{\mathrm{L}}}, \, \ket{-'_{\mathrm{L}}}$ \\
        Fig.~\ref{fig:515_results_loss} & 0.1 & 200 & 3 & $\ket{0_{\mathrm{L}}}, \, \ket{1_{\mathrm{L}}}, \, \ket{+_{\mathrm{L}}}, \, \ket{-_{\mathrm{L}}}, \, \ket{+'_{\mathrm{L}}}, \, \ket{-'_{\mathrm{L}}}$ (50 each)
    \end{tabular}
    \caption{Summary of training hyperparameters.}
    \label{tab:hyperparameters}
\end{table}
Note that the training batch lists the noise-free training states.
These states undergo noise as described in the main text before serving as inputs for the QAEs.
Computational errors are applied by subjecting the training states to the corresponding noise channel, thus only one copy of each training state is contained in the batch.
Erasures, however, are applied probabilistically.
To achieve good statistics we therefore include several copies of each state in the batch when training a collection of QAEs to correct losses of qubits.


\section{Discovered Encodings}\label{appendix:encoding}

As described in Sec.~\ref{section:encoding_discovery} of the main text, certain types of DQNNs can be used to discover logical encodings that optimally protect quantum information from specific kinds of noise.
As an example we trained a collection of 1-4-1 DQNNs to find a logical encoding that perfectly protects states from collective desphasing and furthermore allows for single erasures of qubits to be corrected.
For the training we use a learning rate $\epsilon = 0.1$ and a batch containing $50$ of each of the states $\{ \ket{0}, \, \ket{1}, \, \ket{+}, \, \ket{-}, \, \ket{+'}, \, \ket{-'}$ \}.
Minibatches contain $100$ states and we train for a total number of $150$ epochs.
The logical codespace discovered by the network is spanned by the states
\begin{equation}\label{}
	\begin{aligned}
		\ket{0_{\mathrm{L}}} & = (0.50+0.00i)\ket{0011} + (0.28+0.25i)\ket{0101} - (0.29+0.12i)\ket{0110} \\
		& + (0.33+0.15i)\ket{1001} - (0.30+0.25i)\ket{1010} - (0.31-0.36i)\ket{1100} , \\[3pt]
		\ket{1_{\mathrm{L}}} & = (0.15+0.46i)\ket{0011} + (0.07-0.45i)\ket{0101} - (0.14-0.23i)\ket{0110} \\
		& + (0.10-0.23i)\ket{1001} - (0.05-0.41i)\ket{1010} - (0.48+0.15i)\ket{1100} .
	\end{aligned}
\end{equation}
Note that the vector amplitudes shown here are rounded and components whose squared modulus is smaller than $10^{-3}$ are omitted.
One can clearly see that quantum information is encoded in a DFS.
Logical states $\psiL = \alpha \zeroL + \beta \oneL$ are not altered by the application of a unitary $e^{-i \frac{\alpha}{2} (Z_1 + Z_2 + Z_3 + Z_4)}$ since all computational basis states involved in the logical basis are eigenstates of the ``total magnetization'' $Z_1 + Z_2 + Z_3 + Z_4$ with eigenvalue zero.
Testing the collection of QAEs that results from the trained networks to correct erasures on randomly drawn logical states $\psiL$ we find averaged fidelities between denoised states and corresponding target states that are shown in the first line of Table~\ref{tab:414_fidelities}.
The data indicates that erasures of qubits can be corrected very well.
Given a state $\psiL$, one therefore expects that the marginal state on any single qubit, $\rho_i = \mathrm{Tr}_{\{\mathrm{code \ qubits}\} \setminus i}(\ketbra{\psi_{\mathrm{L}}})$, is maximally mixed and thus the loss of a single qubit does not erase the encoded quantum information.
In the second line of Table~\ref{tab:414_fidelities} we show averaged fidelities of those marginal states w.r.t.~the maximally mixed state.
The fact that they are not exactly equal to one indicates a slight dependence of a marginal state on the coefficients $\alpha$ and $\beta$ of the original logical state.
\begin{table}[H]
    \centering
    \begin{tabular}{c|c|c|c|c|c} 
        & no losses & qubit 1 lost & qubit 2 lost & qubit 3 lost & qubit 4 lost \\ [0.5ex] 
        \hline
        $\langle \mathcal{F} \rangle$ & $1.0000$ & $0.9997$ & $0.9990$ & $0.9997$ & $0.9995$ \\
        $\langle \mathcal{F}(\rho_i, \mathds{1}/2) \rangle$ & $-$ & $0.9999$ & $0.9995$ & $0.9998$ & $0.9997$
    \end{tabular}
    \caption{The first line shows how well a collection of $x$-1-4 QAEs resulting from trained encoding finder networks can denoise random logical states that suffered collective dephasing and qubit erasures. Each network from the collection is tested on $2000$ validation states subjected to collective dephasing of strength $\sigma =1$ and a corresponding loss. The second line shows the potential of the logical encoding to correct single erasure events. This is determined from the averaged fidelity of marginal states on a single qubit, $\rho_i = \mathrm{Tr}_{\{\mathrm{code \ qubits}\} \setminus i}(\ketbra{\psi_{\mathrm{L}}})$, w.r.t.~the maximally mixed state $\mathds{1}/2$.}
    \label{tab:414_fidelities}
\end{table}


\section{Analytical Approach for Noisy QAEs}\label{appendix:noisy_QAE}

In Sec.~\ref{section:internal_noise} of the main text we investigate the quality of a quantum memory that uses an intrinsically noisy QAE for active quantum error correction.
We compare it to a bare physical qubit and to an encoded but uncorrected logical qubit.
Here, we derive analytical expressions for the averaged probability of successful state discrimination that serves as a measure for the quality of such a memory.

A single physical qubit that is subjected to two rounds of bit flip noise of strength $\pidle$, corresponding to two rounds of idling for a time $\tau/2$, suffers a bit flip with probability
\begin{equation}\label{eq:psingle}
    p_{\mathrm{single}} = 2 \pidle (1-\pidle) .
\end{equation}
The probability of successful state discrimination is given by the fidelity of the final state w.r.t. the noise-free initial state $\ket{\psi} = \cos{(\theta/2)} \ket{0} + e^{i \phi} \sin{(\theta/2)} \ket{1}$. Averaging uniformly over all states on the Bloch sphere yields
\begin{equation}\label{eq:fidelity_single}
    \begin{aligned}
        \langle \mathcal{P}_{\mathrm{single}} \rangle = & \iint \frac{d\theta \, d\phi}{4 \pi} \sin(\theta) \Big[ (1-p_{\mathrm{single}}) \braket{\psi} \! \braket{\psi} + p_{\mathrm{single}} \, \expval{X}{\psi} \! \! \expval{X}{\psi} \Big] \\
        = & (1 - p_{\mathrm{single}}) + \frac{1}{3} p_{\mathrm{single}} \\
        = & 1 - \frac{4}{3} \pidle \left( 1 - \pidle \right).
    \end{aligned}
\end{equation}
For an encoded logical qubit that is exposed to two rounds of bit flip noise and then subjected to a perfect round of QEC, the logical error rate is
\begin{equation}
    \begin{aligned}
        p_{\mathrm{uncorr.}} & = p_{\mathrm{single}}^3 + 3 p_{\mathrm{single}}^2 \left( 1 - p_{\mathrm{single}} \right) \\
        & = 12 \pidle^2 (1-\pidle)^4 + 8 \pidle^3 (1-\pidle)^3 + 12 \pidle^4 (1-\pidle)^2 ,
    \end{aligned}
\end{equation}
where $p_{\mathrm{single}}$ is given by Eq.~\eqref{eq:psingle}. Thus, the averaged probability of successful state discrimination reads
\begin{equation}\label{eq:fidelity_logical}
    \begin{aligned}
    \langle \mathcal{P}_{\mathrm{uncorr.}} \rangle = & \iint \frac{d\theta \, d\phi}{4 \pi} \sin(\theta) \Big[ (1-p_{\mathrm{uncorr.}}) \braket{\psi_\mathrm{L}} \! \braket{\psi_\mathrm{L}} + p_{\mathrm{uncorr.}} \, \expval{X_\mathrm{L}}{\psi_\mathrm{L}} \! \! \expval{X_\mathrm{L}}{\psi_\mathrm{L}} \Big] \\
    = & (1 - p_{\mathrm{uncorr.}}) + \frac{1}{3} p_{\mathrm{uncorr.}} \\
    = & 1 - 8 \pidle^2 + \frac{80}{3} \pidle^3 - 40 \pidle^4 + 32 \pidle^5 - \frac{32}{3} \pidle^6.
    \end{aligned}
\end{equation}
\begin{figure}
    \centering
    \includegraphics[width=.45\linewidth]{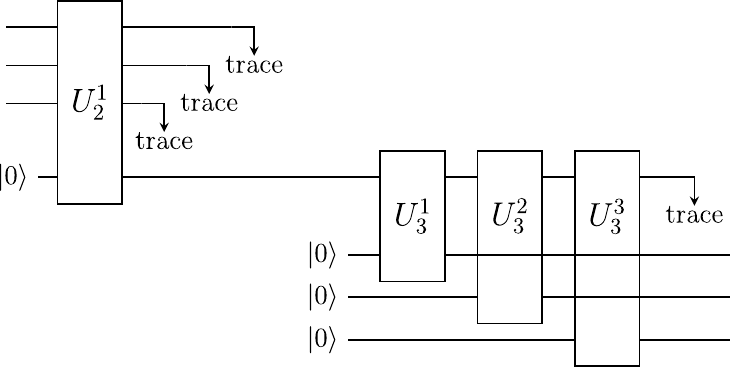}
    \caption{Quantum circuit implementing a 3-1-3 QAE.}
    \label{fig:313_circuit}
\end{figure}
Finally, we consider an intrinsically noisy 3-1-3 QAE that is used for QEC after the first round of bit flip noise.
For weak internal noise, $\pnetwork \ll 1$, output states of the QAE can be well approximated by expanding to linear order in $\pnetwork$.
We consider a hand-constructed QAE that implements the 3-qubit error correction channel stated in Eq.~\eqref{eq:3QC_Krausmap}.
The QAE consists of four unitary matrices, as shown in Fig.~\ref{fig:313_circuit}, where
\begin{figure}[H]
    \centering
    \includegraphics[width=0.18\linewidth]{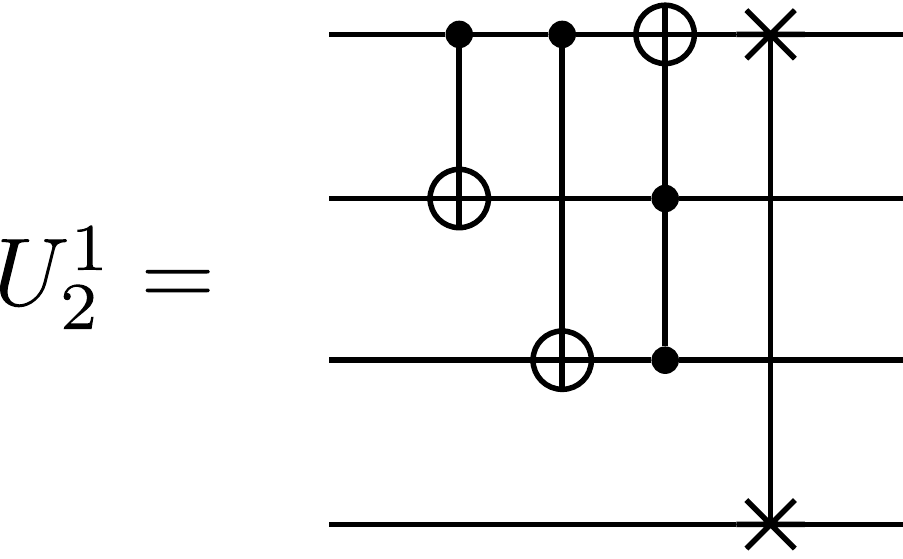}
\end{figure}
\noindent and $U_3^1 = \mathrm{CNOT}$, $U_3^2 = \mathrm{CNOT}$, $U_3^3 = \mathrm{SWAP}$.
Each unitary matrix is followed by a multi-qubit depolarizing channel.
In the absence of internal noise, a logical state corrupted by bit flip noise is mapped to a logical state $\rho_{\mathrm{L}}$ by the QAE.
At the four positions in the circuit where noise is acting, the respective intermediate quantum state remains unaffected with probability $(1-\pnetwork)$ and suffers an error with probability $\pnetwork$.
Expanding an output state of the network to linear order in $\pnetwork$ therefore involves errors at one position in the circuit at most.
For the analysis it turns out to be convenient to write the depolarizing channel acting on $m$ qubits in the form
\begin{equation}\label{}
	\mathcal{N}_{\pnetwork} \left( \rho \right) = \left(1 - \frac{4^m \, \pnetwork}{4^m - 1} \right) \, \rho + \left( \frac{4^m \, \pnetwork}{4^m - 1} \right) \frac{I^{\otimes m}}{2^m} .
\end{equation}
Denoting the output state of the noise-free QAE as $\rho_{\mathrm{L}}$, one can write the output states of the circuit with only one of the four noise channels being present as follows: 
\begin{equation}\label{}
	\rho_{\mathrm{out}}^{\mathrm{noise \, 1}} = \left(1 - \frac{256}{255} \pnetwork \right) \rho_{\mathrm{L}} + \frac{256}{255} \pnetwork \, \frac{\ketbra{000} + \ketbra{111}}{2} ,
\end{equation}
\begin{equation}\label{}
	\rho_{\mathrm{out}}^{\mathrm{noise \, 2}} = \left(1 - \frac{16}{15} \pnetwork \right) \rho_{\mathrm{L}} + \frac{16}{15} \pnetwork \, \frac{\ketbra{000} + \ketbra{011} + \ketbra{100} + \ketbra{111}}{4} ,
\end{equation}
\begin{equation}\label{}
	\rho_{\mathrm{out}}^{\mathrm{noise \, 3}} = \left(1 - \frac{16}{15} \pnetwork \right) \rho_{\mathrm{L}} + \frac{16}{15} \pnetwork \, \big( \bra{00_{23}} \rho_{\mathrm{L}} \ket{00_{23}} + \bra{11_{23}} \rho_{\mathrm{L}} \ket{11_{23}} \big) \otimes \frac{I^{\otimes 2}}{4} ,
\end{equation}
\begin{equation}\label{}
	\rho_{\mathrm{out}}^{\mathrm{noise \, 4}} = \left(1 - \frac{16}{15} \pnetwork \right) \rho_{\mathrm{L}} + \frac{16}{15} \pnetwork \, \big( \bra{0_{3}} \rho_{\mathrm{L}} \ket{0_{3}} + \bra{1_{3}} \rho_{\mathrm{L}} \ket{1_{3}} \big) \otimes \frac{I}{2} .
\end{equation}
Collecting terms yields an approximate expression for a noisy output state of the quantum autoencoder that is linear in $\pnetwork$, resulting from the circuit with all four noise channels present:
\begin{equation}\label{eq:output_noisy}
	\rho_{\mathrm{out}}^{\mathrm{noisy}} \approx \rho_{\mathrm{out}}^{\mathrm{noise \, 1}} + \rho_{\mathrm{out}}^{\mathrm{noise \, 2}} + \rho_{\mathrm{out}}^{\mathrm{noise \, 3}} + \rho_{\mathrm{out}}^{\mathrm{noise \, 4}} - 3 \rho_{\mathrm{L}} .
\end{equation}
To account for the round of bit flip noise before the application of the noisy QAE, we substitute $\rho_{\mathrm{L}} = (1-p_{\mathrm{L}}) \ketbra{\psi_{\mathrm{L}}} + p_{\mathrm{L}} X_{\mathrm{L}} \ketbra{\psi_{\mathrm{L}}} X_{\mathrm{L}}$ into Eq.~\eqref{eq:output_noisy}, where $\pL = \pidle^3 + 3\pidle^2 (1-\pidle)$.
After applying a second round of bit flip noise and performing a perfect round of QEC, we take the fidelity of the resulting state w.r.t.~the original noise-free state $\psiL$ to obtain the probability of successful state discrimination.
Averaging uniformly over all states on the Bloch sphere yields
\begin{equation}\label{eq:fidelity_logical_noisy}
    \begin{aligned}
	    \langle \mathcal{P}_{\mathrm{corr.}} \rangle = & \left(1 - 4 \pidle^2 + \frac{8}{3} \pidle^3 + 12 \pidle^4 - 16 \pidle^5 + \frac{16}{3} \pidle^6 \right) \\
	    & - \pnetwork \bigg(\frac{156}{85} + \frac{8}{15} \pidle - \frac{776}{51} \pidle^2 + \frac{5312}{765} \pidle^3 + \frac{13408}{255} \pidle^4 - \frac{17152}{255} \pidle^5 + \frac{17152}{765} \pidle^6 \bigg) .
	\end{aligned}
\end{equation}
Equating Eq.~\eqref{eq:fidelity_logical_noisy} with Eq.~\eqref{eq:fidelity_single} or Eq.~\eqref{eq:fidelity_logical} and solving for $\pnetwork$ gives rise to analytical expressions for the phase boundaries as shown in Fig.~\ref{fig:phase_diagram} in the main text:
\begin{equation}
	\pnetwork^{\mathrm{crit., \, single}} = \frac{255 \left(\pidle - 4 \pidle^2 + 2 \pidle^3 + 9 \pidle^4 - 12 \pidle^5 + 4 \pidle^6 \right)}{351 + 102 \pidle - 2910 \pidle^2 + 1328 \pidle^3 + 10056 \pidle^4 - 12864 \pidle^5 + 4288 \pidle^6} = \frac{255}{351} \pidle + \mathcal{O}(\pidle^2)
\end{equation}
and
\begin{equation}
	\pnetwork^{\mathrm{crit.,\, logical}} = \frac{765 \left(\pidle^2 - 6 \pidle^3 + 13 \pidle^4 - 12 \pidle^5 + 4 \pidle^6 \right)}{351 + 102 \pidle - 2910 \pidle^2 + 1328 \pidle^3 + 10056 \pidle^4 - 12864 \pidle^5 + 4288 \pidle^6} = \frac{765}{351} \pidle^2 + \mathcal{O}(\pidle^3) .
\end{equation}
%

\end{document}